\documentclass[english,10pt,aps,showpacs]{revtex4}
\usepackage[T1]{fontenc}
\usepackage[utf8]{inputenc}

\usepackage{color}
\usepackage{babel}

\usepackage{array}
\usepackage{longtable}
\usepackage{float}
\usepackage{booktabs}
\usepackage{amsmath}
\usepackage{graphicx}
\usepackage{amssymb}
\usepackage{esint}
\usepackage[unicode=true, 
 bookmarks=true,bookmarksnumbered=false,bookmarksopen=false,
 breaklinks=false,pdfborder={0 0 0},backref=false,colorlinks=false]
 {hyperref}

\hypersetup{pdftitle={CMB in a box: causal structure and
the Fourier-Bessel expansion},
 pdfauthor={L. Raul Abramo, Paulo H. Reimberg & Henrique S. Xavier},
 pdfsubject={cosmology}}

\newcommand{\be}{\begin{equation}}
\newcommand{\ee}{\end{equation}}

\begin{document}

\title{CMB in a box: causal structure and
the Fourier-Bessel expansion}

\author{L. Raul Abramo\footnote{Email: abramo@fma.if.usp.br}}

\affiliation{Instituto de F\'isica, Universidade de S\~ao Paulo, 
CP 66318, 05314-970, S\~ao Paulo, Brazil}

\affiliation{Department of Astrophysical Sciences, Princeton University,
Peyton Hall, Princeton, NJ 08544, USA}

\author{Paulo H. Reimberg and Henrique S. Xavier}

\affiliation{Instituto de F\'isica, Universidade de S\~ao Paulo, 
CP 66318, 05314-970, S\~ao Paulo, Brazil}

\begin{abstract}
This paper makes two points.
First, we show that the line-of-sight solution to cosmic microwave anisotropies in Fourier space, even though formally defined for arbitrarily large wavelengths,
leads to position-space solutions which only depend on the sources of 
anisotropies inside the past light-cone of the observer. This foretold
manifestation of causality in position (real) space
happens order by order in a series expansion in powers of the 
visibility $\gamma = e^{-\mu}$, where $\mu$ is the
optical depth to Thompson scattering.
We show that the contributions of order $\gamma^N$ to
the CMB anisotropies are regulated by spacetime window functions which 
have support only inside the past light-cone of the point of observation.
Second, we show that the Fourier-Bessel expansion of the physical fields
(including the temperature and polarization momenta) is an alternative
to the usual Fourier basis as a framework to compute the anisotropies.
The viability of the Fourier-Bessel series for treating the CMB is a 
consequence of the fact that the visibility function becomes
exponentially small at redshifts $z \gg 10^3$, effectively cutting off the past 
light-cone and introducing a finite radius inside which initial conditions can affect
physical observables measured at our position $\vec{x}=0$ and time $t_0$.
Hence, for each multipole $\ell$ there is a discrete tower of 
momenta $k_{i \, \ell}$ (not a continuum) which can affect physical observables, 
with the smallest momenta being $k_{1 \, \ell} \sim \ell$.
The Fourier-Bessel modes take into account precisely the information 
from the sources
of anisotropies that propagates from the initial value surface to the
point of observation -- no more, no less.
We also show that the physical observables (the temperature and polarization
maps), and hence the angular power spectra, are unaffected by that choice of 
basis. This implies that the Fourier-Bessel expansion is the optimal scheme with which
one can compute CMB anisotropies.
%These results may lead to 
%improvements in codes that compute anisotropies and to 
%simulations of temperature and polarization maps.
\end{abstract}


\maketitle

\section{Introduction}

The cosmic microwave background (CMB) is the earliest, cleanest observation
that reveals what the Universe looked like at the very beginning. A remarkable 
string of observations of the CMB temperature fluctuations over the last 20 years, 
most notably by COBE-DMR \cite{Smoot:1992td} and
WMAP \cite{Bennett:2003bz,Spergel:2003cb,Komatsu:2010fb}, 
has shown that the typical initial conditions of the Universe when it was
under $400.000$ years old can be characterized by extreme 
homogeneity and isotropy, only slightly perturbed by small, 
${\cal{O}}(10^{-5})$ fluctuations with a nearly scale-invariant spectrum. 
More recently, the small degree of polarization that is imprinted on the 
CMB radiation by anisotropic Thompson scattering has also started to 
become detectable \cite{Page:2006hz,Brown:2009uy,Chiang:2009xsa}, 
and may hold the key to unravel the mistery of the birth of our 
Universe -- see, e.g., 
\cite{Dodelson:03,Mukhanov:05,Durrer:2008,Peter:2009zzc}.

The remarkable success of the CMB as probably the most powerful tool in
observational Cosmology can also be traced to the simplicity of its
underlying mechanisms: Thompson scattering and linear perturbation theory.
The basic theory, which is an application of the relativistic radiative transfer 
equations \cite{Chandrasekhar}, was initially developed in 
connection with the CMB by Peebles and Yu \cite{Peebles:1970ag}, and 
the first to write down the full collisional
Boltzmann equations for temperature and polarization 
were Bond and Efstathiou \cite{Bond:1984fp,Bond:1987ub}.

However, even if the main mechanisms driving acoustic oscillations 
in the baryon-photon fluid were basically understood early on,
crucial features such as neutrinos, gravitational waves, spatial 
curvature and the effects of lensing on polarization
remained puzzling.
It was not until the 1990's that the theory reached full maturity, 
when the complete separation of radial and angular modes 
allowed by the use of spin angular momentum eigenfunctions for 
polarization cleared the way for our current understanding of CMB physics 
\cite{Ma:1995ey,Hu:1995en,Hu:1996qs,Seljak:1996gy,Zaldarriaga:1996xe,Kamionkowski:1996zd,Kamionkowski:1996ks,Hu:1997hp,Hu:1997mn}.

One particularly significant step forward was achieved with the 
line-of-sight solution to the collisional Boltzmann equations \cite{Seljak:1996is}. 
The idea is that photons travel along null geodesics,
hence the comoving distance $\Delta x$ between two successive 
collisions is equal to the conformal time interval
$\Delta \eta$ between those collisions (as usual, we assume that the Born 
approximation is valid.) This means that, given a line-of-sight $\hat{n}$, 
we know that a photon detected at time $\eta$ travelling along that 
line-of-sight was at the position $\vec{x}'=\hat{n} (\eta-\eta') $ at time $\eta'$, 
if no collisions occurred between those times. 
The more general case of an ensemble of photons can be easily 
accomodated in this picture, since the probability that a photon 
scatters with free electrons is given in terms of the optical
depth for Thompson scattering, which to a very good approximation is a 
smooth function of time, $\mu(\eta)$.
The final state of the ensemble is then obtained through integration
over time of the sources of temperature and polarization 
anisotropies, appropriately weighted by the optical depth at each time.

The power of the line-of-sight integral solution is that it separates, as much
as it is possible, the (free) propagation of photons from the ultimate 
sources of anisotropies (matter and metric perturbations) -- so, it is 
similar in spirit to a Green's function for the temperature and 
polarization of an ensemble of photons.
However, there is one feature of the generation of anisotropies which
makes it impossible to completely separate the sources and the anisotropies:
anisotropic Thompson scattering is itself a source of both
temperature and polarization, so the process is, in some sense, non-local.

When photons scatter at a given place and time,
the fluctuations in temperature and polarization that are generated as
a consequence of those scatterings depend also on
the quadrupole of the temperature and the polarization of the photons
that were incident at that place and time. Therefore, the fact that 
those incident photons typically propagated to the location of the 
scattering from far away implies that the process is non-local -- hence
the line-of-sight solution is actually a set of integral equations, at least
for the lowest multipoles ($\ell \leq 2$). The higher multipoles 
($\ell \geq 3$), however, can be completely determined from the 
lowest multipoles, which makes the line-of-sight solution a vastly
superior method compared to the usual hierarchy of Boltzmann equations.
Of course, for the low multipoles the integral equations are impractical,
and the preferred method to compute them is to revert back to the hierarchy
of Boltzmann equations, which is then truncated at a relatively low
multipole that is sufficient to accurately compute the multipoles $\ell \leq 2$.

Nevertheless, despite the fact that the generation of anisotropies is a 
non-local mechanism, it is still completely causal: photons propagate along
light cones between scatterings (which are basically instantaneous
within the cosmological timescales.) The main drive behind this work is
to clarify how causality and non-locality are manifested in the generation 
of the CMB. We are interested, in particular, in describing the generation 
of the CMB in position (real) space, since Fourier space can sometimes obfuscate 
the causal nature of the physical mechanisms.

We will show that, in position space, the non-local anisotropies can be resolved 
into explicitly causal pieces by iterating the integral line-of-sight equations. 
The smallness parameter in this perturbative expansion (which is 
reminiscent of the Dyson series of Quantum Electrodynamics 
\cite{Sakurai}) is the {\it total} visibility $\gamma=e^{-\mu}$, where $\mu$ 
is the optical depth to Thompson scattering.
The first term in that series corresponds to the last scattering of the
photons before they were detected; the second term corresponds to 
the last two scatterings before detection; and so on.

\begin{figure}[t]
%\begin{centering}
\includegraphics[scale=0.75]{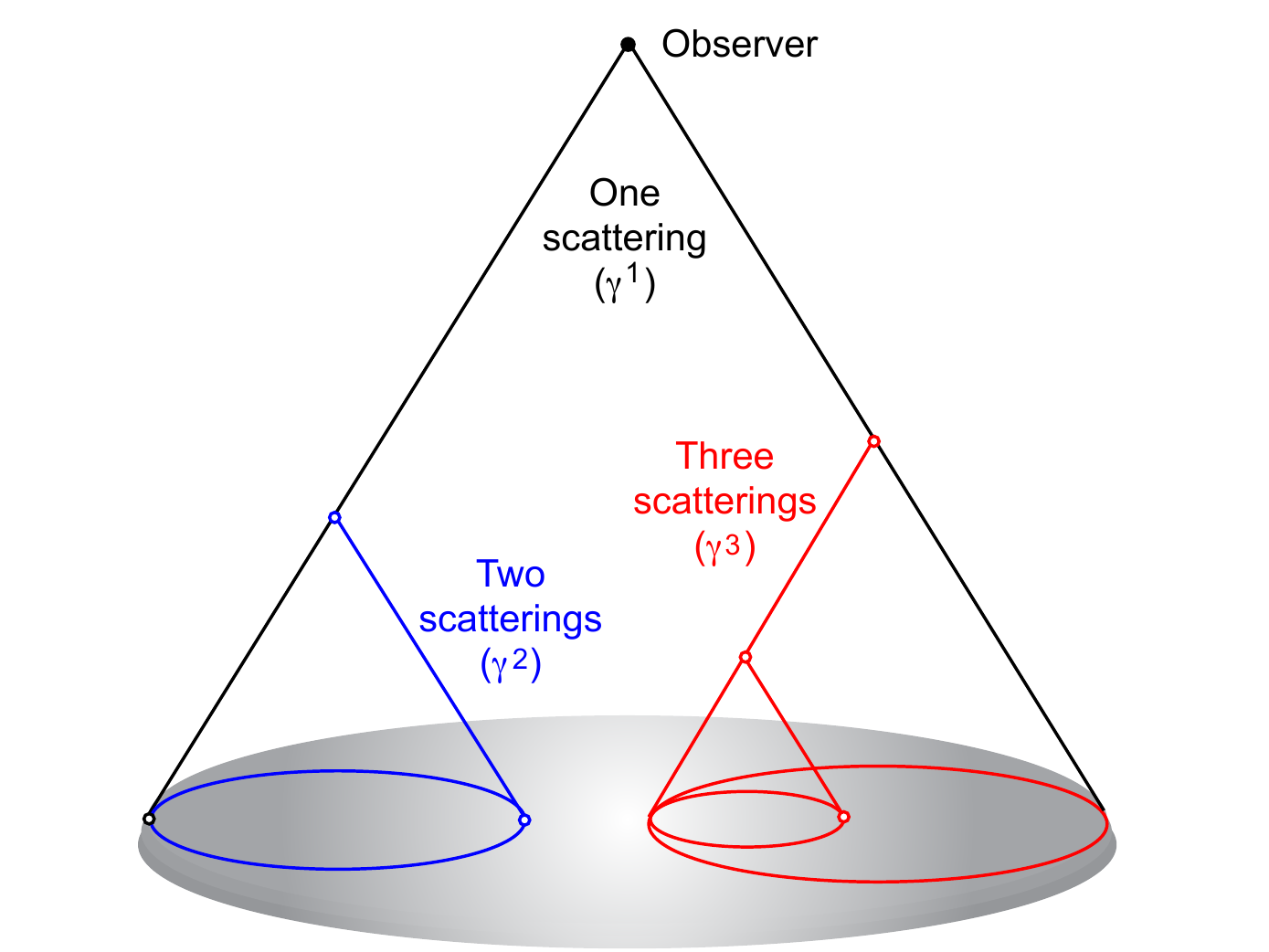}
%\par\end{centering}

\caption{Past light-cones for photons experiencing one, two and three 
scatterings since the time of decoupling from matter (denoted as the 
base of the cone.) The terms corresponding to these interactions
are respectively of order $\gamma^1$, $\gamma^2$ and $\gamma^3$, where
$\gamma=e^{-\mu}$ is the visibility.}

\label{LightCones}
\end{figure}

In this series over the number of scatterings, causality is 
manifested at each order in terms of
radial integrals corresponding to spacetime window functions
which are only non-zero inside the past light-cone (PLC) of
the observation point. The spacetime dependence of the 
$\gamma^N$-order term is given in terms of integrals of $N+1$ spherical
Bessel functions, which we demonstrate to have support only
inside the PLC. The first term of the series, ${\cal{O}}(\gamma^1)$, 
corresponds simply to the Sachs-Wolfe (SW), Doppler and 
Integrated Sachs-Wolfe (ISW) contributions to the temperature fluctuations 
from the last scattering \cite{1967ApJ...147...73S}, and evidently it 
is only nonzero 
at the surface of the light cone. The second-order terms for temperature
and polarization carry the memory of the last two scatterings, and
can be non-zero anywhere inside the volume of the PLC 
(not only on its surface). Fig. \ref{LightCones} illustrates the structure of the light 
cones for one, two and three scatterings.
Hence, we have shown that the line-of-sight integrals are actually
{\it retarded} Green's functions for the temperature and polarization, and
we found analytical expressions for them in position space.

The second result of this paper follows from this causal structure.
The expansion in successive scatterings (or, equivalently, in orders of the visibility 
function) makes it clear that what matters for the CMB observables 
are not simply the source fields
and the anisotropies as functions of position and time, $f(\vec{x},\eta)$, but 
the fields multiplied by powers of the visibility, $\gamma(\eta)^N f(\vec{x},\eta)$.
Since the visibility vanishes for very early times
(say, $z_i >> 10^3$), for all practical purposes that have to do with
the CMB, our PLC is cut-off at $\eta_i=\eta(z_i)$, and all the sources
are effectively zero outside the radius corresponding to that time (which 
in a typical flat $\Lambda$CDM model 
is approximately $r_i \approx 5 H_0^{-1}$.)

The fact that the physical fields are exponentially suppressed at the
boundaries of the PLC means that the best basis for expanding the
fields, as well as the anisotropies, is not the Fourier basis, which is 
most convenient for plane waves in $\mathbb{R}^3$,
but the Fourier-Bessel basis \cite{Watson}, 
which expands functions $f(\vec{x})$ inside a
3D sphere of finite radius $R$ into spherical harmonics $Y_{\ell m} (\hat{x})$
and spherical Bessel functions $j_\ell(k_{i \ell} x)$, with $k_{i \ell}R$
being the $i^{th}$ root of $j_\ell$. Both sets of basis functions 
are eigenvectors of the Laplacian operator in flat space, with eigenvalues
${-\vec{k}}^2$ or $-k_{i \ell}^2$, but they differ most significantly in that
the Fourier-Bessel series establishes a discrete
tower of momenta for each multipole $\ell$, so that the smallest momentum
mode is $k_{1 \ell} \sim \ell/r_i$. Therefore, there is a clearly defined 
minimal mode that can contribute to CMB observables,
and the modes above it are all discretized. 
%In other words: the window functions become 
%trivial in the Fourier-Bessel basis. 
The drawback is that, if the underlying fields are Gaussian in nature, 
the Fourier-Bessel
modes, as opposed to the Fourier modes, are not statistically 
independent, so their covariance matrix is not diagonal. 
However, dynamics is one thing, statistics is another, and we will 
show how these issues can be separated so that we can easily 
recover angular power spectra which have precisely the same
statistical properties of the power spectra computed
in Fourier space.

The first part of this work takes an approach which 
is similar to that used to make constrained simulations of CMB 
temperature and polarization maps by 
Liguori, Matarrese and Moscardini \cite{Liguori:2003mb} and by
Komatsu, Spergel and Wandelt \cite{Komatsu:2003iq} 
-- see also \cite{Cornish:2003db,Liguori:2007sj,Elsner:2009md}.
However, while those simulations make use of several approximation
schemes in order to produce numerically viable codes, our expressions 
are exact and analytical. In particular, our results show that the transfer 
functions of \cite{Liguori:2003mb,Komatsu:2003iq} have 
an invariant (or geometrical) piece which is described by our spacetime window
functions, so that the transfer functions can be obtained by integration of these 
window functions over time with some visibility function.
Another work close in spirit to ours was done by Bashinsky and
Bertshinger \cite{Bashinsky:2000uh,Bashinsky:2002vx}, who calculated 
the Green's function for the evolution of linear cosmological perturbations 
in position space but did not compute
the Green's functions for the anisotropies.

A quick note on our conventions: the Fourier transform of a function
$f(\vec{x})$ is $f(\vec{k}) = (2\pi)^{-3/2}\int d^3 x\, e^{-i \vec{k} \cdot \vec{x}}
f(\vec{x})$; the perturbed Friedman-Robertson-Walker metric is
$ds^2 = a^2(\eta) [-(1+2\Phi)d\eta^2 + (1-2\Psi) d\vec{x}^2 ]$; 
and the relationship between the rotation matrices and the spin
spherical harmonics is $D^\ell_{m, -s} (\alpha,\beta,\gamma \rightarrow 0)
= \sqrt{4\pi/(2\ell+1)} \, \, {}_{s} Y_{\ell m}^* (\beta,\alpha)$, so that 
${}_s Y^*_{\ell m} = (-1)^{m-s} \, {}_{-s} Y_{\ell, -m}$.
Finally, everything we will say in this paper concerns scalar 
(spin-0) density perturbations -- we will consider lensing 
and gravitational waves in future work.

%%%%%%%%%%%%%%%%%%%%%

\section{CMB anisotropies in a nutshell}

The main channel of electron-photon interactions during 
recombination is elastic Thompson scattering. The likelihood of
a photon interacting with free electrons between times 
$\eta_1$ and $\eta_2$ can be determined in terms of the optical
depth for Thompson scattering:
\be
\label{OPD}
\mu(\eta_1,\eta_2) = \int^{\eta_2}_{\eta_1} d\eta \, a(\eta) \, 
\sigma_T \, n_e (\eta) X_e(\eta) \; ,
\ee
where $a(\eta)$ is the Friedman-Robertson-Walker scale factor, 
$ \sigma_T = 6.65 \times 10^{-25}$ cm$^2$ is 
the Thompson cross section,
$n_e$ is the total number density of electrons, and $X_e$ is the ionized fraction.
The probability per unit time that a photon observed at $\eta_2$ had last 
interacted at time $\eta_1$ is called the {\it visibility function}:
\be
\label{VIF}
g(\eta_1,\eta_2) = \frac{d}{d\eta_1} e^{- \mu (\eta_1,\eta_2)} \; ,
\ee
where $g(\eta_1,\eta_2)$ is a positive-definite function, normalized to unity
if the limits are taken such that $\mu \rightarrow 0$ and $\mu \rightarrow \infty$ at
late and early times, respectively. 
It is customary to define the
visibility function today simply by $g(\eta) = g(\eta,\eta_0)$.
In this work we will also define the {\it total visibility} (or
unscattered fraction) as:
\be
\gamma (\eta_1,\eta_2) = e^{- \mu (\eta_1,\eta_2)} \; .
\label{GAM}
\ee
Notice that the {\it total} visibility is the probability that a photon will {\it not}
scatter after $\eta_1$ before it is detected at $\eta_2$, and in fact
we defined it such that it is related to the visibility {\it function} by
$g(\eta_1,\eta_2) = d/d\eta_1 [\gamma (\eta_1,\eta_2)]$.

At very early times, the optical depth is extremely large, so the Universe
is effectively opaque and the visibility vanishes exponentially. 
Photons and electrons are interacting
so often that the photons can spend enough time in some small
region so that inelastic processes lead to thermalization with the
electrons and, by extension, with the baryons. Hence, at very
early times (much before recombination) the photon distribution 
$\Theta(\vec{x},\eta;\hat{l})=\Delta T/T$ 
was essentially in equilibrium with baryonic matter, and only the monopole
$\theta_0 (\vec{x},\eta) = \int d^2 \hat{l}/(4\pi) \Theta (\vec{x},\eta;\hat{l})
= \delta_\gamma/4$ was significant (here $\delta_\gamma$ is the density
contrast of photons.)

However, as soon as recombination starts the optical depth plummets,
scatterings between photons and electrons become sparser, and as a result 
the radiation incident on any given scattering source can be increasingly anisotropic, since photons arriving to a scattering source from 
distant over- or underdense regions have different equilibrium temperatures. 
Thomson scattering with anisotropic radiation
then generates polarization (and vice-versa), and the process becomes quite
intricated.

Let us assume for a moment that the photons decoupled from matter instantly,
at some time $\eta_{R}$, and propagated freely from that time
down to our detectors at $ \vec{x}=0$,  $\eta_0$ . Then direct 
integration of the geodesic equation 
(with the help of the Born approximation) tells us that the 
temperature from a given line-of-sight $\hat{l}$ in fact reflects the
density, gravitational potential and velocities at the position 
$\vec{x}_l = (\eta_0 - \eta_R) \, \hat{l}$ and time $\eta_R$, as well as any 
time-varying gravitational potentials along that light-cone 
\cite{1967ApJ...147...73S}:
\be
\label{GEO}
\Theta(\vec{x}=\vec{0},\eta_0;\hat{l}) = 
\left[ \theta_0 + \Phi + \hat{l} \cdot \vec{\nabla} V_b \right] (\vec{x}_l,\eta_R)
+ \int^{\eta_0}_{\eta_R} d\eta (\Phi'+\Psi') (\vec{x}_l,\eta_R) \; ,
\ee
where $V_b$ is the baryon velocity potential. This equation shows
that the primary (and ultimate) sources of anisotropies are the
inhomogeneites in the matter and metric fields of the Universe 
($\theta_0$, $\Phi$, $\Psi$ and $V_b$.)

Now let's relax the assumption of instant recombination, but still 
require that the photons did not scatter again after decoupling. 
This is the same as saying that the visibility function is not 
assumed to be proportional to a delta-function $\delta(\eta-\eta_R)$ anymore, 
but is still a positive, normalized function, highly peaked at the time of recombination.
In that case the photons will carry the average temperature of 
the location where they last scattered, along with the local 
gravitational potential $\Phi$ and baryon velocity $V_b$, and this signal
will be affected by the time-varying metric perturbations 
only after that last scattering. 
Considering that the probability that a photon will
scatter between some time $\eta'$ and the some time $\eta$ 
is given by the visibility function $g(\eta',\eta)$, but the probability
that they will {\it not} scatter anymore after the time $\eta'$ is given
by the {\it total} visibility $\gamma(\eta',\eta)$, the line-of-sight 
solution becomes:
\be
\label{LOS}
\Theta^{(1)} (\vec{0},\eta;\hat{l}) = \int_{0}^{\eta} \, d\eta' \,
\left\{
g(\eta',\eta)
\left[ \theta_0 + \Phi + \hat{l} \cdot \vec{\nabla} V_b \right] (\vec{x}_l,\eta')
+ \gamma(\eta',\eta)
(\Phi'+\Psi') (\vec{x}_l,\eta') \right\}
\; ,
\ee
where now $\vec{x}_l =  \Delta\eta \, \hat{l}$, with $\Delta\eta = (\eta - \eta')$.
The superscript $1$ is used to denote that this contribution is linear with 
in the total visibility $\gamma$ (as well as the visibility function, 
$g=d\gamma/d\eta$).

Equation (\ref{LOS}) tells us that, in a first approximation, to obtain the
temperature anisotropies one should simply average the sources 
over the PLC $\vec{x}_l$, with weights given either 
by the visibility function (for the SW and Doppler terms) or by the 
total visibility (for the ISW term.) This approximation would correspond to the 
``one scattering'' diagram of Fig. 1.

The next level of complexity leads to polarization. Let's assume that
the approximation above is still true for the temperature, but that
photons can scatter a second time after decoupling. 
Then, the incident radiation
at the location of that scattering will in general be
anisotropic, simply because of the inhomogeneities in the Universe at the
time of decoupling. If that incident radiation has a quadrupole, then Thompson
scattering will excite the linear polarization degrees of freedom of
the Stokes parameters $Q$ and $U$. Clearly, then, polarization
is of at least second order in the visibility, since it enters once
when the photons first decouple from the matter, and then a second
time when the anisotropic ensemble of photons scatter off free
electrons, generating the polarization.

The assumptions above are too simplistic, of course: as we go back
in time the number of scatterings per Hubble time rise steeply, which
means that the problem that must be solved is one of 
sucessive scatterings of a polarized, inhomogeneous and anisotropic
temperature distribution which is, moreover, coupled to baryons and 
dark matter. 

The result of taking into account the anisotropy and polarization
of the incident radiation in Thompson scattering leads to corrections 
to Eq. (\ref{LOS}) and to the generation of polarization
\cite{Bond:1987ub,1996AnPhy.246...49K,Hu:1997hp,Dodelson:03,Straumann:2005mz,Durrer:2008,Peter:2009zzc}. 
The polarization at any given point in space and time 
is best given in terms of the (dimensionless) spin +2 eigenstate 
combination:
\be
\label{PQU}
\frac{Q+iU}{4 I}  \equiv P(\vec{x},\eta;\hat{l}) = \sqrt{\frac{3}{40 \pi} }
\int_{0}^{\eta} \, d\eta' \, g(\eta',\eta)
\left[ \theta_2 (\vec{x}',\eta') - \sqrt{6} \, p_2  (\vec{x}',\eta') \right]
\; ,
\ee
where $\theta_2$ and $p_2$ are the quadrupole of the temperature and
of the polarization, which we will define in more detail below.
To these equations we should naturally add the perturbed Einstein equations, 
as well as the continuity and Euler equations for baryons, dark matter 
and neutrinos -- see, e.g., \cite{Ma:1995ey,Dodelson:03,Peter:2009zzc}.

It should be evident from the symmetries of the problem 
that it is natural to break this system 
of equations into spherical coordinates with respect to the 
lines-of-sight $\hat{l}$ around an observer at the origin. 
For temperature, which is a scalar under rotations, the spherical 
harmonic decomposition reads:
\be
\label{TYL}
\Theta(\vec{x},\eta;\hat{l}) = \sum_{\ell m} \Theta_{\ell m} (\vec{x},\eta)
Y_{\ell m} (\hat{l}) \; .
\ee
Polarization, on the other hand, is such that the Stokes parameters
$Q$ and $U$ change sign if we perform a rotation of $\pi$ over
the line-of-sight, which means that they are 
components of a spin-2 field. In fact, the
complex combination in Eq. (\ref{PQU}) was chosen such that it is
a spin +2 eigenstate. Hence, polarization in this form can be
expanded in terms of the spin +2 eigenfunctions, or spin +2 spherical
harmonics \cite{Kamionkowski:1996ks,Hu:1997hp}:
\be
\label{PYL}
P(\vec{x},\eta;\hat{l}) = \sum_{\ell \geq 2, m} P_{\ell m} (\vec{x},\eta)
\, {}_2 Y_{\ell m} (\hat{l}) \; .
\ee

In Fourier space the dependence on $\hat{l}$ can be
easily isolated, since
$e^{i \vec{k} \cdot \vec{x}' }= e^{i \vec{k} \cdot \vec{x} } e^{i \Delta\eta
\vec{k} \cdot \hat{l} }$. We then employ Rayleigh's expansion, 
$e^{i \vec{k} \cdot \vec{x}} = 4\pi  \sum_{\ell m} i^\ell j_\ell(kx) 
Y^*_{\ell m}(\hat{k}) Y_{\ell m}(\hat{x})$, and
the line-of-sight integrals determining anisotropies
can be written in the form \cite{Seljak:1996is,Hu:1997hp}:
\begin{eqnarray}
\label{TMO}
\Theta_{\ell m} (\vec{x},\eta) &=&
4\pi i^\ell \int \frac{d^3 k}{(2\pi)^{3/2}}
e^{i \vec{k} \cdot \vec{x}} \, Y^*_{\ell m} (\hat{k}) \, \theta_\ell (\vec{k},\eta) \; ,
\\ 
\label{PMO}
P_{\ell m} (\vec{x},\eta) &=&
4\pi i^\ell \int \frac{d^3 k}{(2\pi)^{3/2}}
e^{i \vec{k} \cdot \vec{x}} \, Y^*_{\ell m} (\hat{k}) \, p_\ell (\vec{k},\eta) \; ,
\end{eqnarray}
where the temperature and polarization momenta, $\theta_\ell$ and
$p_\ell$, are derived from the the geodesic equation for photons
in the presence of Thompson scattering:
\begin{eqnarray}
\label{TPO}
\theta_\ell (\vec{k},\eta) &=& \theta^{(1)}_\ell (\vec{k},\eta) + 
\frac{1}{4} \int_{0}^{\eta} \, d\eta' \, g(\eta',\eta) 
\left[ \theta_2 (\vec{k},\eta') - \sqrt{6} \, p_2  (\vec{k},\eta') \right]
\left[ 1+ 3 \frac{ \partial^2}{\partial (k \Delta\eta)^2 } \right]
j_\ell(k \Delta \eta) \; ,
\\
\label{TP1}
\theta^{(1)}_\ell (\vec{k},\eta) &=&
\int_{0}^{\eta} \, d\eta' \,
\left\{ g(\eta',\eta)
\left[ \theta_0 (\vec{k},\eta') + \Phi(\vec{k},\eta') 
+ V_b(\vec{k},\eta') \frac{\partial}{\partial\eta} \right] 
+ \gamma(\eta',\eta)
(\Phi'+\Psi') (\vec{k},\eta') \right\} 
j_\ell (k\Delta \eta) \; ,
\\
\label{PPO}
p_\ell (\vec{k},\eta) &=& - \frac{3}{4} \sqrt{\frac{(\ell+2)!}{(\ell-2)!}} 
\int_{0}^{\eta} \, d\eta' \, g(\eta',\eta) 
\left[ \theta_2 (\vec{k},\eta') - \sqrt{6} \, p_2  (\vec{k},\eta') \right]
\frac{j_\ell(k \Delta \eta)}{(k \Delta \eta)^2} \; .
\end{eqnarray}
For the derivation of the polarization term for the temperature
quadrupole, including the radial function, see also \cite{Abramo:2006gp}.

Therefore, in this form the structure of the interactions is more clear
than in the hierarchy of Boltzmann equations, but at the price of
stating the problem in terms of integro-differential equations.
This complexity is just apparent, though, since all higher-order
source terms have angular momenta $\ell \leq 2$, and if we solve 
for the low ones, all the higher multipoles can be computed with the 
help of the integrals above \cite{Seljak:1996is}.

%%%%%%%%%%%%%%%%%%
\section{CMB in position space and causality}
\label{SecCau}

The integral equations (\ref{TPO})-(\ref{PPO}) should be solved, 
in principle, for all momenta $\vec{k}$ so that the temperature and
polarization anisotropies in Eqs. (\ref{TMO})-(\ref{PMO}) can be computed.
This includes the modes with $k << H_0$, which correspond
to arbitrarily large wavelengths and can contribute to the zero mode
of the fluctuations. Now, does this mean that perturbations
outside the horizon can contribute anything to the CMB that we observer? 
In fact they don't:
we will show that the temperature and polarization of the CMB which
are observed at any spacetime point $(\vec{x},\eta)$
only include information from inside the PLC of that point.
This statement is true at each order in the visibility  $\gamma$, 
and for each spherical mode $(\ell, m)$.

The first step to recover the causal structure which underlies the 
temperature and polarization is to go from Fourier space
back to position space. The most direct way to go back to position space
without relinquishing the spherical harmonic decomposition is to
use the fact that the Fourier- and position-space  harmonics 
are simply related by a Hankel transform:
\begin{eqnarray}
\label{FLM}
f(\vec{x}) = \sum_{LM} f_{LM}(x) Y_{LM} (\hat{x}) 
\quad &,& \quad 
f(\vec{k}) = \sum_{LM} f_{LM}(k) Y_{LM} (\hat{k}) \; , 
\\
\label{HTR}
f_{LM}(x) = \sqrt{\frac{2}{\pi}} i^L
\int_0^\infty dk \, k^2 \, j_L (kx) \,  f_{LM}(k) \quad &,& \quad
f_{LM}(k) = \sqrt{\frac{2}{\pi}} (-i)^L
\int_0^\infty dx \, x^2 \, j_L (kx) \, f_{LM}(x) \; .
\end{eqnarray}
If $f(\vec{x})$ is a real function, then the harmonic coefficients
obey the conjugation relations $f_{\ell m}^*(r) = (-1)^m f_{\ell , -m} (r)$
in position space and $f_{\ell m}^*(r) = (-1)^{\ell+m} f_{\ell , -m} (r)$ in
Fourier space.

The relations above between $f_{LM}(x)$ and $f_{LM}(k)$ are 
quite remarkable: they tell us
that the spherical harmonic phases do not mix at all. 
This is a consequence \cite{1996MNRAS.278...73H} of the fact that angular
momentum is the same operator in position and in Fourier space,
${\bf L} = i \vec{x} \times \vec{\partial}_{x} = i \vec{k} \times \vec{\partial}_{k} $ .
It is worth noting that apparently this technique were first used in 
Cosmology in connection with redshift space distortions -- see, e.g., 
\cite{Heavens:1994iq,1996MNRAS.278...73H,Hamilton:1997zq}. 
In connection with the CMB, the spherical decomposition 
has been used in simulations \cite{Liguori:2003mb,Komatsu:2003iq}, 
and as a tool to study polarization from clusters of 
galaxies by \cite{Bunn:2006mp,Abramo:2006gp}.

\subsection{$\gamma^1 $ term: $\Theta^{(1)}$ in position space}
\label{SubGamma1}

Now we can easily substitute the sources in terms
of this spherical harmonic decomposition into $\theta^{(1)}_\ell$. 
In the following subsections we 
show how this prescription can be extended to the 
remaining terms in the expressions (\ref{TPO}) and (\ref{PPO}).

Let's then express the monopole $\theta_0$, 
newtonian potential $\Phi$ and baryon velocity potential $V_b$ in terms
of spherical harmonics $Y_{LM}(\hat{x})$, and use them to compute
the temperature anisotropies at our location (assuming that we 
occupy the origin of the spherical coordinate system, at $\vec{x}=0$), 
to first order in the visibility. Substituting the spherical
harmonic decomposition in $\vec{x}$ into Eq. (\ref{TP1}) for $\theta^{(1)}$,
inserting that expression in Eq. (\ref{TYL}) and integrating 
over $d^2\hat{k}$ (which makes $L=\ell$ and $M=m$), leads to:
\begin{eqnarray}
\label{T1R}
\Theta^{(1)}_{\ell m} (\vec{0},\eta) =
\frac{2}{\pi} \int_0^\infty dk \, k^2 \int_0^\eta d\eta' \int_0^\infty dx \, x^2 \, 
S_{\ell m} (x,\eta,\eta') \, j_\ell (k\Delta \eta) j_\ell (k x) \; ,
\end{eqnarray}
where we have collected the sources in the term:
\be
\label{SOU}
S_{\ell m} (x,\eta,\eta') = g(\eta',\eta)
\left[ \theta_{0,\ell m} (x,\eta') + \Phi_{\ell m}(x,\eta') 
- V_{b,\ell m}(x,\eta') \frac{\partial}{\partial\eta'} \right] 
+ \gamma(\eta',\eta)
(\Phi'+\Psi')_{\ell m} (x,\eta') \; .
\ee
Notice that we have defined the sources in an unusual way, including
a factor of the visibility function for the Sachs-Wolfe and Doppler terms,
and a factor of the total visibility for the integrated Sachs-Wolfe term.
The matter and metric fields are actually just functions of $(x,\eta')$, but
we include the extra dependence on $\eta$ that comes from the visibility
into the definition of the source term in order to simplify the notations.

Now the integral over $k$ in Eq. (\ref{T1R}) can be computed, and in 
fact that happens to be exactly the orthogonality condition for 
spherical Bessel functions:
\be
\label{OBF}
\int_0^\infty dk \, k^2 \, j_\ell(kx) \, j_\ell(kx') = \frac{\pi}{2} x^{-2} 
\delta (x-x') \; .
\ee
This implies that Eq. (\ref{T1R}) can be simplified to:
\begin{eqnarray}
\label{T1R2}
\Theta^{(1)}_{\ell m} (\vec{0},\eta) &=&
\int_0^\eta d\eta' \, \int_0^\infty dx \, S_{\ell m} (x,\eta,\eta') \, \delta(x-\Delta\eta)
\\ \nonumber
&=&\int_0^\eta d\eta' \, \left\{ g(\eta',\eta)
\left[ \theta_{0,\ell m} (\Delta \eta,\eta') + \Phi_{\ell m}(\Delta \eta,\eta') 
+ V_{b,\ell m}'(\Delta \eta,\eta') \right] 
+ \gamma(\eta',\eta)
(\Phi'+\Psi')_{\ell m} (\Delta \eta,\eta') \right\}  \; ,
\end{eqnarray}
which is just the harmonic decomposition of  Eq. (\ref{LOS}). Notice that
in Eq. (\ref{LOS}) $\hat{x}=\hat{l}$, so the gradient in the Doppler term 
can be written as a time derivative, which after integration by
parts with the derivative of the delta-function becomes the derivative of
the baryon velocity in the expression above. This warm-up exercise is
useful to check that all sources which contribute to temperature 
anisotropies at this level come from the light-cone (its surface, in this 
case), which here appears explicitly as $\delta(x-\Delta\eta)$.

Another remarkable fact, which already shows up in this lowest-order
approximation but which is true to all orders, is that the phases
$(\ell,m)$ of the CMB observables are the same as the phases
of the sources. The only conditions for this to hold are, first, that 
we keep to linear perturbation theory, and second, that the optical 
depth is a function of time only. We will see next that this holds true 
to higher orders in the total visibility $\gamma$.

\subsection{$\gamma^2$ terms in position space}
\label{SubGamma2}

The expressions (\ref{TPO})-(\ref{PPO}) are integral 
equations for the temperature and polarization momenta $\theta_\ell$
and $p_\ell$. We can iterate these equations and organize the series
into powers of the total visibility, similarly to what is done
for the Dyson series of Quantum Electrodynamics \cite{Sakurai} -- except
that the fields in the integral equations for the CMB are coupled to a 
set of ordinary differential equations (the Einstein, continuity and Euler 
equations for metric and matter perturbations.)

In the previous subsection we computed the first term of this series,
which is of order $\gamma^1$ (since the source term is itself linear in
$\gamma$). Consider now the next terms of this series, 
which are of order $\gamma^2$.

For polarization, we have that, to order $\gamma^2$, the only term which
contributes is the temperature quadrupole to order $\gamma^1$. 
By substituting Eq. (\ref{TP1}) into Eq. (\ref{PPO}) and 
expressing the sources in terms of their spherical harmonic
decompositions, like was done in the previous subsection, we obtain
after some algebra:
\begin{eqnarray}
\label{P2R}
P_{\ell m}^{(2)} (\vec{0},\eta) &=& - \frac{3}{2\pi} \sqrt{\frac{(\ell+2)!}{(\ell-2)!}} 
\int_0^\eta d\eta' \, g(\eta',\eta) 
\int_0^{\eta'} d\eta'' \, \int_0^\infty dx \, S_{\ell m} (x,\eta',\eta'')  
\\ \nonumber
& \times & 
\int_0^\infty dk \, (k x)^2 \, j_\ell (k x)   \, 
\frac{j_\ell (k \Delta \eta)}{(k \Delta \eta)^2} \, 
j_2 (k \Delta\eta')  \; ,
\end{eqnarray}
where $\Delta \eta'= \eta'- \eta''$. A simplified version of
Eq. (\ref{P2R}) was first written in this form by 
\cite{Abramo:2006gp}, for the case of the polarization
from a galaxy cluster -- where the visibility function after decoupling
can be thought of being proportional to a Dirac $\delta$-function,
$g_c(\eta,\vec{x}) = \mu_c \, \delta(\vec{x}-\hat{n} \,\Delta\eta )$,
where $\mu_c$ is a cluster's optical depth and $\hat{n}$ is
the line-of-sight to the cluster.

\begin{figure}[t]
\begin{centering}
\includegraphics[scale=0.75]{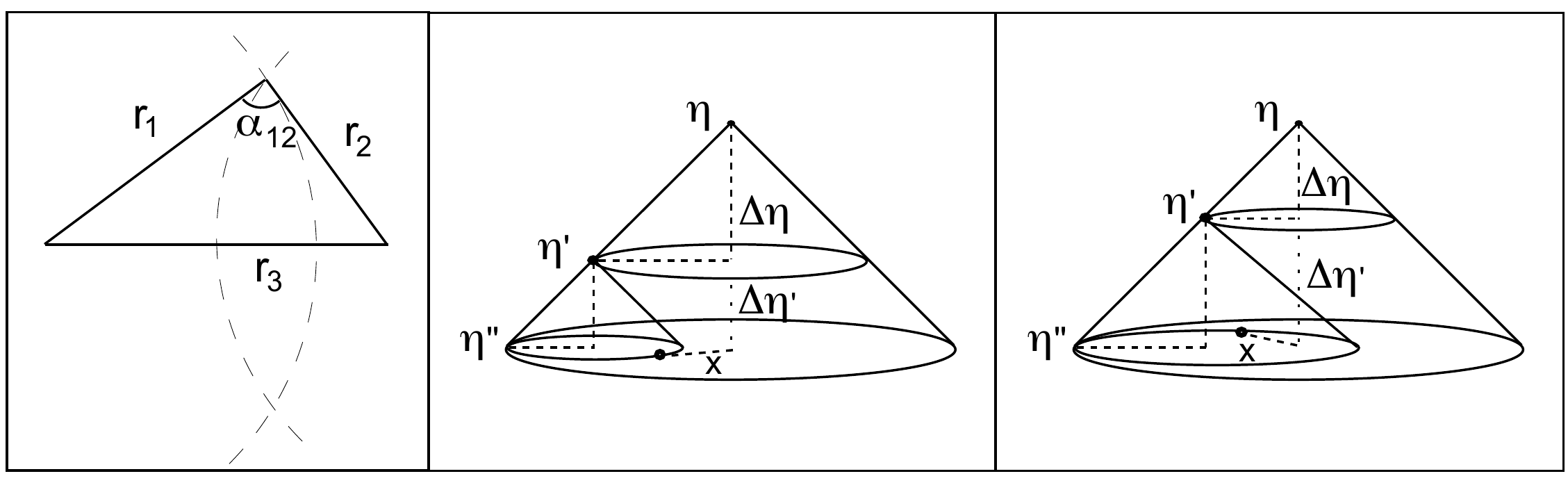}
\par\end{centering}
\caption{Left panel: if the triangle of sides $r_1$, $r_2$ and $r_3$ exists, then
the window function $W^3_\ell$ is non-zero. Middle and right panels: 
spacetime diagrams for the scatterings, where time runs up, and space
(radial coordinates) runs horizontally from the center to the sides. 
The diagrams show
sources which decoupled at $\eta''$, then scatter off free electrons at time 
$\eta'$, and end up contributing to the polarization 
$P^{(2)}_{\ell m}$ at time $(\eta)$.
The minimal and maximal values of $x$ for which the
sources can contribute to polarization
are given by $x_{min} = |\Delta \eta - \Delta \eta'|$ and 
$x_{max}=\Delta \eta + \Delta \eta' = \eta-\eta''$.}
\label{Triangles}
\end{figure}

Physically, this contribution to polarization corresponds to the
photons that decoupled at some radius $x$ and time $\eta''$,
are scattered at time $\eta'$, and then end up as a polarized 
beam at time $\eta$. The geometry is shown in Fig. 2.
We will show now that the integral over $k$ at the end of Eq. (\ref{P2R})
has exactly that meaning: it vanishes unless the distances
$x$, $\Delta \eta$ and $\Delta \eta'$ form a triangle, and
such a triangle does not exist unless the sources are inside
the PLC of the last scattering point, and the point of last scattering lies
inside the PLC of the observation point. 
We will give a general expression for integrals such as this in Appendix 
\ref{ApInts}, but here is the result \cite{Watson}:
\begin{eqnarray}
\label{IK3}
W^3_\ell(r_1,r_2;r_3) &=& \frac{r_1^2}{r_2^2}
\int_0^\infty dk \, j_\ell (k r_1) \, j_\ell (k r_2)   \, j_2 (k r_3) 
\\ \nonumber
& = & \frac{\pi}{4} \frac{r_1^3 }{r_2 r_3^3} P_\ell^{(-2)}(\cos\alpha_{12}) 
\sin^2\alpha_{12} 
\; ,
\end{eqnarray}
where:
\be
\label{CAL}
\cos\alpha_{12}= \frac{r_1^2 + r_2^2 - r_3^2}{2 r_1 r_2}
\ee
is the cosine of the angle between the sides $r_1$ and $r_2$ in the triangle of
sides $r_1$, $r_2$ and $r_3$. 
The window function $W^3_\ell$ is zero if that triangle does
not exist, which means that it is nonzero only if the following set of 
conditions are satisfied:
\be
\label{IN3}
r_1 \leq r_2 + r_3 \quad , 
\quad r_2 \leq  r_3 + r_1 \quad , 
\quad r_3 \leq r_1+r_2 \; ,
\ee
or, equivalently, $|r_1-r_2| \leq r_3 \leq r_1+r_2$ 
-- see the left panel of Fig. \ref{Triangles}. This window
function was first computed in Ref. \cite{Abramo:2006gp}. We have 
plotted some cuts of those window functions in the Appendix, 
Fig \ref{W3}.

Hence the momentum integral in Eq. (\ref{T2R}) is 
$W^3_\ell (x,\Delta\eta;\Delta\eta')$, which means that it 
is a {\it spacetime window function} that vanishes unless
the inequalities above are satisfied.
What this result implies to our Eq. (\ref{P2R}) is that a 
source at radius $x$ and time $\eta''$ 
can only contribute to $P^{(2)}_{\ell m}(\eta)$ 
if $|\Delta \eta - \Delta \eta'| \leq x \leq \Delta \eta + \Delta \eta'=\eta-\eta''$. 
As the right panel of Fig. 2 shows, this spacetime window function
limits the contributions of the sources to the PLCs of the last
scatterings, and the scatterings themselves to the PLC of the 
observation point. Obviously, this means that the sources that contribute to 
$P^{(2)}_{\ell m}$ must also be inside the PLC of the observation point --
which in this case is $\eta-\eta''$.

Let us summarize this result for the lowest-order contribution to polarization
in position space:
\be
\label{P2R2}
P_{\ell m}^{(2)} (\vec{0},\eta) = - \frac{3}{2\pi} \sqrt{\frac{(\ell+2)!}{(\ell-2)!}}
\int_0^\eta d\eta' \, g(\eta',\eta) 
\int_0^{\eta'} d\eta'' \, \int_0^{\eta-\eta''} dx \, 
 S_{\ell m} (x,\eta',\eta'')  \,  W_\ell^3 (x,\Delta\eta; \Delta\eta') 
\ee
Notice that, among other restrictions, the window function above 
cuts off the spatial integral at radius $x$ at $\eta-\eta''$, which is 
where the PLC of the observation point lies at time $\eta''$.

In \cite{Abramo:2006gp} Eq. (\ref{P2R2}) was used as the starting point to
show how to invert the polarization data from galaxy clusters in order 
to reconstruct the three-dimensional map of the sources at the time of decoupling. That constitutes the solution (to lowest order in visibility) to 
a conjecture by Kamionkowsky and Loeb about how to get around 
cosmic variance using cluster polarization
\cite{Kamionkowski:1997na}.

Therefore, just as happened in the previous subsection with 
$\Theta^{(1)}_{\ell m}$, the integral over $k$ in effect guarantees
that the sources will only be taken into account if the scattering
processes happen on the light cones.
As before, the integral over $x$ in Eq. (\ref{P2R}) is 
cut-off at $x_{max} = \eta-\eta''$, 
eliminating sources which lie outside the PLC of the observation point
$(\vec{0},\eta)$. But now there is another feature: since
$x_{min} = | \Delta\eta-\Delta\eta' |$,
sources which were too close to the observation point at
time $\eta''$ also cannot contribute to the CMB at time $\eta$
if last scattering happened at time $\eta'$ --
see the middle and right panels of Fig. \ref{Triangles}. 
This additional constraint on the
volume of the PLC which is integrated applies for times of 
last scattering ($\eta'$) which are both close and far from 
the observation time $\eta$.

Before we turn to the order $\gamma^3$ terms, we write down the
order $\gamma^2$ contribution to the temperature anisotropies, which
comes from inserting $\theta^{(1)}_2$ into Eq. (\ref{TPO}). 
After a calculation very similar to the one 
done above for polarization, we obtain that:
 \begin{eqnarray}
\label{T2R}
\Theta_{\ell m}^{(2)} (\vec{0},\eta) &=& \frac{1}{4} 
\int_0^\eta d\eta' \, g(\eta',\eta) 
\int_0^{\eta'} d\eta'' \, 
\int_0^\infty dx \, 
S_{\ell m} (x,\eta',\eta'') 
\\ \nonumber
& \times & 
\int_0^\infty dk \, (kx)^2 \, j_\ell (k x) \, 
\left[ 1 + 3 \frac{\partial^2}{\partial (k \Delta \eta)^2} \right] j_\ell (k \Delta\eta)
 \, j_2 (k \Delta \eta') \; .
\end{eqnarray}
The integral over $k$ on the last line of the previous equation 
can be recast in terms of $W^3_\ell$ if we 
use the recursion relations for the derivatives of spherical Bessel functions:
\be
\label{REC}
z^{\mp \ell} \frac{d}{dz} \left[ z^{\pm \ell} j_\ell(z) \right] = \pm j_{\ell \mp 1} (z) \; .
\ee
The momentum integral then becomes:
$$
(x\Delta \eta)^{-\ell-1} \frac{d}{dx} \frac{d}{d\Delta\eta} 
\left[ x^{\ell-1} \Delta \eta^{\ell+3} W^3_{\ell+1} (x,\Delta\eta; \Delta\eta') \right]
+ 3 \frac{\partial^2}{\partial \Delta \eta^2} 
\left[ \frac{\Delta\eta^2}{x^2} W^3_\ell (x,\Delta\eta; \Delta\eta') \right]
\; ,
$$
which shows that, just as happened for polarization, 
the contribution of order $\gamma^2$ to the temperature is also
modulated by the same spacetime window function.

\subsection{$\gamma^3$ and higher-order terms in position space}

\begin{figure}[t]
\begin{centering}
\includegraphics[scale=0.75]{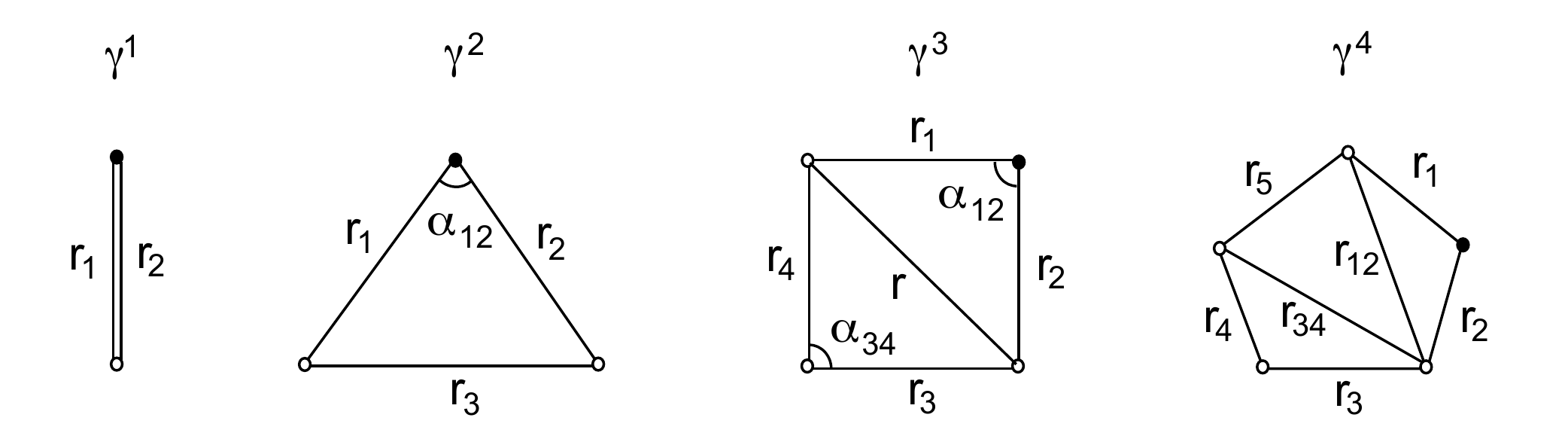}
\par\end{centering}
\caption{Diagrammatic representation of the series in the visibility.
In all diagrams, the filled dot corresponds to the point and time
of observation of the CMB temperature and polarization.
The $\gamma^1$ term (the two lines beginning and ending at
the same points) corresponds to a window function  
$\delta(r_1-r_2)$. 
The $\gamma^2$ and $\gamma^3$ terms (respectively
the triangle and 4-side polygon) 
correspond to the window functions 
$W^3_\ell$ of Eq. (\ref{IK3}) and
$W^4_\ell$ of Eq. (\ref{IK4}).
For all diagrams the window functions vanish
unless the sides are such that the polygon can be closed. The main
physical implication
is that sources (which by default are located at $r_1$ with
respect to the observation point) 
outside the PLCs of the last scatterings do not 
contribute to the CMB. In particular, sources outside the PLC
of the point of observation are thrown out of the integration of
the physical observables. See the Appendix for a full discussion
of these geometrical properties.}
\label{Polygons}
\end{figure}

The iteration process is trivial, but the higher-order terms 
become lengthy. We present the result for the
simplest $\gamma^3$ term that contributes to polarization:
\begin{eqnarray}
\label{P3R}
P_{\ell m}^{(3)} (\vec{0},\eta) &=& - \frac{27}{2\pi} \sqrt{\frac{(\ell+2)!}{(\ell-2)!}}
\int_0^\eta d\eta' \, g(\eta',\eta) 
\int_0^{\eta'} d\eta'' \, g(\eta'',\eta') 
\int_0^{\eta''} d\eta''' \, 
\int_0^\infty dx \, S_{\ell m} (x,\eta'',\eta''') \, 
\\ \nonumber
& \times & 
\int_0^\infty dk \, (kx)^2 \, j_\ell (k x) \, 
\frac{j_\ell (k \Delta\eta)}{(k \Delta\eta)^2}   \, 
\frac{j_2 (k \Delta\eta')}{(k \Delta\eta')^2}  \,
j_2 (k \Delta \eta'')
   \; ,
\end{eqnarray}
where $\Delta \eta''= \eta''- \eta'''$.
We will show in Appendix \ref{ApInts} that the integral over $k$
can be resolved into: 
\begin{eqnarray}
\label{IK4}
W^4_\ell(r_1,r_2;r_3,r_4) &=& 
\frac{r_1^2}{r_2^2 r_3^2}
\int_0^\infty dk \, k^{-2} \, j_\ell (k r_1) \, j_\ell (k r_2)   \, j_2 (k r_3) \, j_2 (k r_4) 
\\ \nonumber
& = & 
\frac12 
\int d(\cos\alpha_{34}) \frac{r_4^2}{r^2}
P_2^{(-2)}(\cos\alpha_{34}) \sin^2\alpha_{34} 
\times \frac{\pi}{4} \frac{r_1^3}{r_2 r^3} 
P_\ell^{(-2)}(\cos\alpha_{12}) \sin^2\alpha_{12} 
\; ,
\end{eqnarray}
where now $r$ is defined as the common side of two triangles,
of sides $(r_1,r_2,r)$ and $(r_3,r_4,r)$, such that
$r^2 = r_1^2 + r_2^2 - 2 r_1 r_2 \cos\alpha_{12} = 
r_3^2 + r_4^2 - 2 r_3 r_4 \cos\alpha_{34}$ -- see Fig. \ref{Polygons}.
It is trivial to show that all the remaining terms of order $\gamma^3$ 
that contribute to the CMB temperature and polarization can be written 
in terms of the window function $W^4_\ell$ by using the recursion 
relations of spherical Bessel functions.

\begin{figure}[t]
\begin{centering}
\includegraphics[scale=0.75]{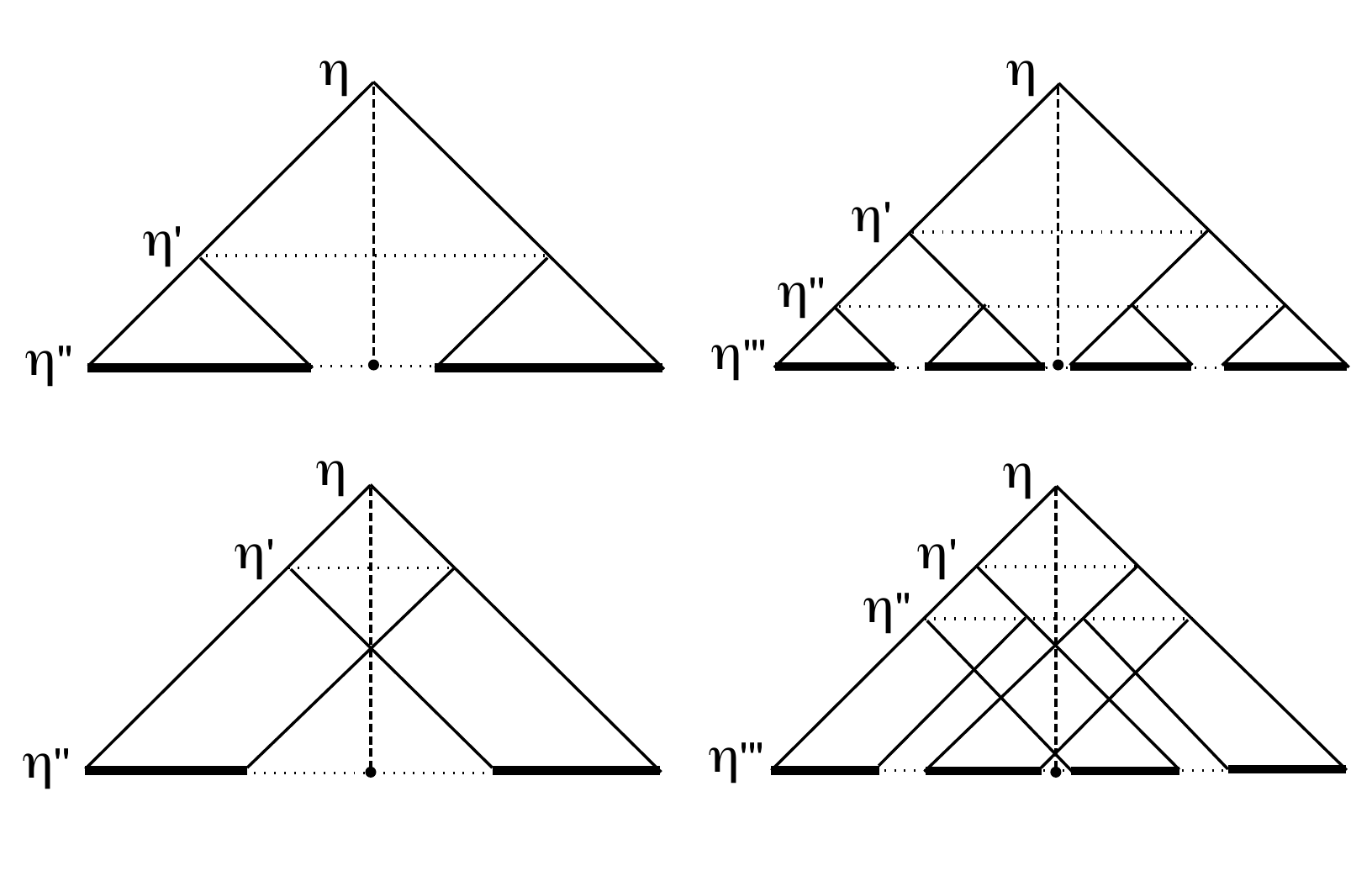}
\par\end{centering}
\caption{Structure of the successive PLCs for two (left diagrams) and 
three (right diagrams) scatterings. The values of $x$ for which the 
window functions $W_\ell^3(x,\eta;\eta')$ and 
$W_\ell^4 (x,\eta;\eta',\eta'')$ are non-zero are indicated as the thick 
lines at the bases of the cones. Here the radius
$x$ should be measured from the origin of the spatial coordinates, which
in these spacetime diagrams lie along the dashed lines. Notice that the region
in the vicinity of the PLC of the observation point (the outermost PLC) always
contributes to the integral, the origin is always excluded, 
and intermediate regions may or may not
contribute to the CMB observables.}
\label{PLCs}
\end{figure}

The window function vanishes unless all four sides can form a (flat) polygon,
so inequalities similar to those found for $W^3_\ell$ apply here
as well:
\begin{eqnarray}
\label{IN4}
r_1 \leq r_2 + r_3 + r_4 \quad &,& 
\quad r_2 \leq  r_3 + r_4 + r_1 \; , \\ \nonumber
r_3 \leq r_4 + r_1+r_2 \quad &,&
\quad r_4 \leq r_1 + r_2 + r_3 \; .
\end{eqnarray}
These conditions imply once again that the sources can only contribute
to the observable at time $\eta$ if each successive PLC lies inside the
previous PLC, all the way from the observation point back to the sources.
The constraints on the positions of the sources are now more 
complicated than the case of two scatterings, but the diagrams in 
Fig. \ref{PLCs} show which values of $x$ (measured from the origin, at
the middle of the bases of the cones) are allowed by the window
functions. In three spatial dimensions, these ranges of radii correspond
to concentric spherical shells -- in the case of two scatterings, there is
one spherical shell; in the case of three scatterings, two spherical
shells; and so on. The outermost spherical shell always includes the edge of
the PLC of the observation point.

It is easy to recover $W^3_\ell$ from $W^4_\ell$ by taking $r_3 \rightarrow 0$.
In fact, we can also recover the orthogonality condition for the spherical
Bessel functions, Eq. (\ref{OBF}), from $W^3_\ell$ by taking $r_3 \rightarrow 0$.
This can be made by noticing that, in $W^3_\ell$ the limit $r_3 \rightarrow 0$
leads to $r_2=r_1$, and in $W^4_\ell$ the limit $r_3 \rightarrow 0$
leads to $r_4=r$. Then, using the expansion of the Bessel functions for small
arguments, $j_2(z) \approx z^2/15 + {\cal{O}}(z^4)$, we obtain that:
\begin{eqnarray}
\label{LI3}
\lim_{r_3 \rightarrow 0} W^3_\ell(r_1,r_2;r_3) &=& \frac{r_3^2}{15} \times 
\frac{\pi}{2} r_1^{-2} \delta (r_1-r_2) \; ,
\\
\label{LI4}
\lim_{r_3 \rightarrow 0} W^4_\ell(r_1,r_2;r_3,r_4) &=& \frac{1}{15} \times 
 W^3_\ell(r_1,r_2;r_3) \; .
\end{eqnarray}
The first identity just shows that the PLC delta-function is the spacetime
window function for two scatterings (order $\gamma^1$.)
The second identity can also be verified by noticing that in
the $r_3 \rightarrow 0$ limit, $\cos\alpha_{34}$ is unscontrained,
so we can integrate out the dependence on that angle, which gives a
factor of $2/15$.

The procedures outlined above can be extended once again to the next order in 
the visibility, $\gamma^4$. The geometrical interpretation is given by the 
rightmost diagram in Fig. \ref{Polygons}.
Again, we see the role played by the spacetime window functions,
of regulating the volume of the PLC in each scattering so that the 
information from the sources is propagated causally all the way to 
the observer. Notice also that by taking $r_3 \rightarrow 0$ we
recover $W^4_\ell$ -- in fact, all the spacetime window functions
obey the relation:
$$
\lim_{r_3 \rightarrow 0} W^n_\ell (r_1,r_2;r_3 , \ldots , r_n)=
\frac{1}{15} W^{n-1}_\ell (r_1,r_2;r_4 , \ldots , r_{n})
$$

\subsection{Interpretation of the series in $\gamma$}

In this Section we have shown, first, that the contribution of order $\gamma^1$ 
from sources at positions $x$ and times of last scattering $\eta'$ 
to the temperature which is observed at time $\eta$ is modulated by a
delta-function, $\delta(x-\Delta\eta)$ -- and this is nothing but the surface
of the PLC of the observation point, $x=\Delta\eta = \eta -\eta'$. 
Then we showed that the contributions of order
$\gamma^2$ to temperature and polarization are modulated by 
the spacetime window function
$W^3_\ell(x,\Delta\eta;\Delta\eta')$, where now $\Delta \eta$ denotes
the interval between observation and the last scattering, 
and $\Delta\eta'$ denotes the interval between that last 
scattering and the second-to-last scattering. 
We presented the simplest contribution of order 
$\gamma^3$, which sums over sources modulated by the
window function $W^4_\ell$, and indicated how the order 
$\gamma^4$ term is also modulated by a window function $W^5_\ell$. 
Notice that we have been labeling each term of the series in terms 
of $\gamma$, which in this notation stands both for the 
{\it total} visibility $\gamma=e^{-\mu}$ 
and for the visibility {\it function} $g=d\gamma/d\eta$.
However, even though the total visibility is always smaller than unity, 
the visibility function is highly peaked at the time of recombination 
and during reionization, so it would be incorrect to characterize our series 
over visibility as a perturbative series -- it is rather more like an 
asymptotic series.

The series over visibility has the diagrammatic representation shown in Fig. 
\ref{Polygons}. Perhaps it is not so surprising that the
spherical Bessel functions play a key role in regulating the
volume inside the PLC, since they are associated with the matrix
elements of $E_3$ -- the Euclidean group in three dimensions
\cite{Talman}. $E_3$ is a non-compact group consisting 
of the set of transformations under which distances are invariant, 
which may explain why the spherical Bessel functions modulate
the (invariant) distances on each equal-time hypersurface.

Another check on our results is the fact that all the window functions are real.
Since the sources $S(\vec{x},\eta,\eta')$ are themselves real, this 
means in particular that $P_{\ell m}^* = (-1)^{m}  P_{\ell, - m}$ -- i.e.,
the polarization that is generated from scalar perturbations is made
up purely of $E$-modes. If we had included gravity waves or lensing, these 
window functions would have acquired an imaginary piece as well, which 
would have ``magnetic'' (instead of ``electric'') parity and therefore 
generate $B$-modes.

The spacetime window functions to any order in $\gamma$
vanish unless each scattering lies in the PLC of the following
scattering, all the way from the sources to the observation point.
The order $\gamma^N$ contributions
to CMB observables hold the information from the $N$-th 
last scatterings, and are modulated by the window function
$W^{N+1}_{\ell} (x,\Delta\eta;\Delta \eta_1, \ldots , \Delta \eta_{N-1})$.
In terms of the
diagrams of Fig. \ref{Polygons}, there are always two spherical 
Bessel functions of order $\ell$, corresponding to the source (x)
and the last scattering before observation ($\Delta \eta$), 
and $N-1$ spherical Bessel functions of order $2$, corresponding 
to the $N-1$ intermediate scatterings for which the 
quadrupoles of the temperature and polarization served as sources 
in the iterations of the line-of-sight integrals. The computation of 
these spacetime window functions, as well as the restrictions 
they impose on the sources that can contribute to the CMB
(in particular the fact that they all vanish outside the PLC of
the observation point) are shown in Appendix \ref{ApInts}.

Therefore, in position space the line-of-sight integral equations have
a simple interpretation in terms of successive Thompson scatterings 
happening along the successive PLCs. These expressions should
still be coupled to the Einstein, continuity and Euler equations
through the temperature dipole and quadrupole, but those are
local equations so their causal structure is trivial
(nevertheless, the Green's function in position space for the
cosmological matter and metric perturbations can also reveal
very interesting features \cite{Bashinsky:2000uh,Bashinsky:2002vx}.)

Going back to the hierarchy of Boltzmann equations, 
which are obtained directly from the line-of sight integrals in Fourier space,
there is one full hierarchy which holds separately for each mode $\{k,\ell,m\}$.
Hence, what we have shown is an explicit demonstration that the 
line-of-sight formalism can also be seen as the solution to an initial
values problem, where the initial conditions only need to
be specified inside the PLC of the observer, at some initial time
when the visibility was small enough that the series converges quickly. 
In other words: we have explicitly shown that the line-of-sight integrals 
in Fourier space are essentially the Fourier transform of a retarded 
Green's function for the CMB observables. These retarded Green
function for anisotropies in position space are expressed at each order in 
the series over visibility through spacetime window functions, which
are ultimately the objects responsible for enforcing causality and regulating
which inhomogeneities are able to affect anisotropies, and how.

The results of this Section have an interesting connection with 
schemes to simulate constrained maps of CMB temperature
and polarization 
\cite{Liguori:2003mb,Komatsu:2003iq,Cornish:2003db,Liguori:2007sj,Elsner:2009md}.
These simulations are extremely important to predict the types and
levels of non-gaussianity in the CMB which are generated at
the very early Universe, since distinct inflationary
models can be differentiated on that basis \cite{Maldacena:2002vr}.
In those simulations, the temperature and polarization transfer functions, 
which transform the inhomogeneities (the sources) into anisotropies,
are computed numerically assuming some
visibility function. Moreover, some approximations
are typically made, such as considering only the curvature perturbation in the
source term. Our expressions, on the other hand, are exact and 
analytical, but it is not immediately clear how (or if) they could be used to 
facilitate a simulation. However, our results show that the transfer 
functions of \cite{Liguori:2003mb,Komatsu:2003iq} are made up of 
invariant (or geometrical) pieces which can be factored from the
purely time-dependent visibility function. These invariant parts are 
given by our spacetime window functions, which do not 
depend on the cosmological scenario or on the history of recombination. 
The transfer functions relevant for the simulations can, therefore, be 
obtained by integrating our spacetime window functions over time with 
some visibility function.

%%%%%%%%%%%%%%%%%%%%%%%%%%%%%%%%%

\section{CMB with the Fourier-Bessel expansion}

We have seen that the line-of-sight integrals for the CMB temperature 
and polarization, when framed in position space, lead to 
spacetime window functions that constrain the positions of the sources 
$S(x)$ that are eventually integrated over. 
These constraints, valid at each order
in the visibility, imply in particular that the sources 
must all lie inside the PLC of the observer that measures the CMB.
But this is not the only constraint that is relevant for this problem.

The physics of recombination is such that the visibility
is exponentially small for $z \gg 10^3$. Since 
all the terms of the series over visibility are linear on the
sources, the fact that they are always
multiplied by powers of the visibility implies that the 
sources which are relevant in the line-of-sight integrals 
are those that lie inside the PLC of the observer at 
some time late enough that the visibility is non-vanishing.
So, the sources which are actually summed into the CMB
anisotropies are not simply the sources in the PLC of
the observer, but the sources in the PLC of the
observer at times such that the
visibility is not totally negligible. Since for a typical flat $\Lambda$CDM
cosmology the lookback distance
from today to those early times saturates near $R \sim 5 H_0^{-1}$
for $z \gg 10^3$, it makes little difference whether we choose
that initial instant (when $\gamma \rightarrow 0$ and $g \rightarrow 0$)
to be $z=10^4$, $10^5$ or $10^{10}$.

This means that the integration of the sources is not
simply limited to the PLC of the observer, as implied by the window 
functions $W^N_\ell$, but that the relevant spacetime
volume is that of the PLC, cut-off at some initial time $\eta_i$
such that $\gamma(\eta,\eta_i)$ and $g(\eta,\eta_i)$ are
sufficiently small. 
Since the unperturbed spacetime is symmetric around the source, 
there is in effect a maximal radius away
from the observer, $R=\eta-\eta_i$, such that outside
that radius, the sources are effectively zero by virtue of the powers
of $\gamma$ that multiply them at each order in the series.
Notice that we could even choose $R$ to be bigger than this lookback
time, but that would unnecessarily include sources which are
eventually discarded in the integration of the physical observables.

Hence, we can fix some boundary $R$ and set all
fields to zero at that boundary and beyond: the result of solving
the initial values problem through the line-of-sight integrals
will be exactly the same, order by order in the series over
the visibility.

If that is the case, then we should ask what would be the best
way to represent the sources, considering that they are
zero at and above some radius $R$ from the origin. The appropriate
expansion in that case is clearly the Fourier-Bessel series \cite{Watson},
for which the fields are expanded in spherical harmonics
and spherical Bessel functions, but instead of the continuum
of momenta that appears in Fourier space, Eq. (\ref{HTR}), 
the modes are discretized: they are given by the roots of the 
spherical Bessel functions. A function that obeys Dirichlet
boundary conditions at $r=R$ is expanded as:
\begin{eqnarray}
\label{FBY}
f(\vec{x}) = \sum_{\ell m} f_{\ell m} (x) \, Y_{\ell m} (\hat{x}) 
= \sum_{\ell m} \sum_{i=1}^\infty f_{i \ell m} \, 
j_\ell (k_{i \ell} x) \,
Y_{\ell m} (\hat{x}) 
\; ,
\end{eqnarray}
where the last sum is over the $i$-th root of $j_\ell$, so that for all 
$ i$'s the Bessel functions vanish at the boundary, 
$j_\ell(k_{i\ell}R) =0$.
The coefficients $f_{i \ell m}$ can be obtained by using the 
orthogonality relation of the Fourier-Bessel basis:
\be
\label{OFB}
\int_0^1 dz \, z^2 \, j_\ell(q_{i\ell} z) \, j_\ell(q_{j\ell} z) =
\frac{1}{2} \, \left[ j_{\ell+1}(q_{i\ell}) \right]^2 \, \delta_{ij} \; ,
\ee
where $q_{i\ell}$ is the $i$-th root of $j_\ell(z)$.
This expression leads immediately to:
\be
\label{IFB}
f_{i\ell m} = \frac{2 R^{-3}}{j_{\ell+1}^2 (k_{i\ell}R)}
\int_0^R dx \, x^2 \, j_\ell(k_{i \ell} x) \, f_{\ell m} (x) \; ,
\ee
where $k_{i\ell} = q_{i\ell}/R$.

The most important feature of the Fourier-Bessel series is that its 
basis functions, like the plane waves of the Fourier expansion, are
eigenvectors of the Laplacian operator in flat space, 
$\nabla^2  \, j_\ell (k_{i \ell} x)  \, Y_{\ell m} (\hat{x})  \,= \, -  \,k_{i \ell}^2 \,
j_\ell (k_{i \ell} x)  \, Y_{\ell m} (\hat{x}) $. This means that the Einstein,
continuity and Euler equations for matter are exactly the same as
in the usual Fourier expansion -- except that the momenta $k_{i\ell}$
are now discretized. Moereover, the lowest eigenmode of
the Fourier-Bessel series that can contribute for a given multipole 
$\ell$ is $k_{1\ell} \sim \ell/R$.

\subsection{Fourier-Bessel modes of CMB observables}

Now we can go back to the CMB and translate the equations
and methods presented in the previous Sections to the
Fourier-Bessel basis. The most important difference between
the two expansions can be grasped by comparing Eqs. (\ref{HTR}), 
(\ref{FBY}) and (\ref{IFB}). Basically, when going from the Fourier
basis to the Fourier-Bessel basis, the angular dependence
is still expressed in terms of spherical harmonics, but
the radial coordinate is expressed by a sum, not an integral:
\be
\label{FFB}
\sqrt{\frac{2}{\pi}} \, i^\ell \, \int_0^\infty \, dk \, k^2 \; 
\rightarrow R^{-3} \sum_{i} \; ,
\ee
which just tells us how to go from the continuum of momenta
appropriate for fields in $\mathbb{R}^3$ to the discrete tower 
of momenta $k_{i\ell}$ that encapsulates all the information for
fields limited to the finite volume inside a sphere of radius R.

In particular, this means that now the CMB temperature and polarization
are given in terms of the momenta not by Eqs. (\ref{TMO}) - (\ref{PMO}), 
but by:
\begin{eqnarray}
\label{TFB}
\Theta_{\ell m} (\eta) &=& \sum_{i} \Theta_{i \ell m} (\eta) \; ,
\\
\label{PFB}
P_{\ell m} (\eta) &=& \sum_{i} P_{i \ell m} (\eta) \; .
\end{eqnarray}
In these sums $\Theta_{i \ell m} \equiv \theta_{\ell, i \ell m}$ and 
$P_{i \ell m}\equiv p_{\ell, i \ell m} $ 
are given by the solutions of integral equations analogous to 
Eqs. (\ref{TPO})-(\ref{PPO}), which for each multipole $L$ 
and for each mode $\{ i \ell m \}$ read:
\begin{eqnarray}
\label{TMB}
\theta_{L, i \ell m} (\eta) &=& \theta^{(1)}_{L,i \ell m} (\eta) + 
\frac{1}{4} \int_{0}^{\eta} \, d\eta' \, g(\eta',\eta) 
\left[ \theta_{2,i\ell m} (\eta') - \sqrt{6} \, p_{2,i\ell m}  (\eta') \right]
\left[ 1+ 3 \frac{ \partial^2}{\partial (k_{i\ell} \Delta\eta)^2 } \right]
j_L (k_{i\ell} \Delta \eta) \; ,
\\
\label{T1B}
\theta^{(1)}_{L,i \ell m} (\eta) &=&
\int_{0}^{\eta} \, d\eta' \,
S_{i\ell m}(\eta,\eta') \, j_L (k_{i\ell} \Delta \eta) \; ,
\\
\label{PMB}
p_{L,i \ell m} (\eta) &=& - \frac{3}{4} \sqrt{\frac{(\ell+2)!}{(\ell-2)!}} 
\int_{0}^{\eta} \, d\eta' \, g(\eta',\eta) 
\left[ \theta_{2,i\ell m} (\eta') - \sqrt{6} \, p_{2,i\ell m}  (\eta') \right]
\frac{j_L(k_{i\ell} \Delta \eta)}{(k_{i\ell} \Delta \eta)^2} \; ,
\end{eqnarray}
and where the sources in Eq. (\ref{T1B}), which
were defined in Eq. (\ref{SOU}),
have been expanded in the Fourier-Bessel basis as well.
Notice that only the generalized modes $\{ L; i , \ell, m \}$ with
$L\leq2$ really need to be calculated from the integral equations
(or, equivalently, from the associated Boltzmann equations), since all 
the higher
modes ($L\geq 3$) can be computed from the former, and 
only the pieces $L=\ell$ actually get summed into the CMB observables,
Eqs. (\ref{TFB})-(\ref{PFB}). This is just a restatement of the
fact that the sources
of anisotropies are the matter and metric inhomogeneities, plus
the dipole and quadrupole of temperature and polarization -- which,
again, underpins the vast superiority of the line-of-sight formalism
compared to the full hierarchy of Boltzmann equations.

An important check of consistency is to reobtain the temperature
and polarization anisotropies in position space that were derived in
Section III. To first order in $\gamma$, the temperature anisotropies
are given by:
\begin{eqnarray}
\label{TFBR}
\Theta_{\ell m}^{(1)} (\eta) &=&
\sum_{i} \theta_{\ell, i \ell m}^{(1)}
\\ \nonumber
&=& 
\int_{0}^{\eta} \, d\eta' \,
\int_0^R dx \, x^2 \, 
\sum_{i} \,
\frac{2 R^{-3}}{j_{\ell+1}^2 (k_{i\ell}R)} \,
j_\ell(k_{i \ell} x) \, S_{\ell m} (x,\eta,\eta')
\, j_\ell (k_{i\ell} \Delta \eta) \; ,
\end{eqnarray}
where the source term $S_{\ell m} (x,\eta,\eta')$ was defined in Eq. (\ref{SOU}).
But now the infinite sum in Eq. (\ref{TFBR}) can be resolved through the 
use of the orthogonality of the Fourier-Bessel basis in target space 
(actually, this is a completeness relation -- see \cite{Watson}, Cap. XVIII), 
which is just the Fourier-Bessel counterpart of Eq. (\ref{OBF}):
\be
\label{OBB}
\sum_{i=1}^\infty \frac{j_\ell (k_{i \ell} \, r_1) 
\, j_\ell (k_{i \ell} \,  r_2)}{j_{\ell+1}^2 (k_{i \ell} \,  R)}
= \frac12 \, R^3 \, r_1^{-2} \, \delta(r_1-r_2) \; .
\ee
This identity leads then automatically to Eq. (\ref{T1R2}), which shows
that to order $\gamma^1$ the Fourier and Fourier-Bessel descriptions
are identical.

To order $\gamma^2$ it is less obvious that one obtains the same
anisotropies as related to the position-space fields, but it is true nevertheless.
To see that, take the simplest term -- the contribution to polarization that
comes from the order $\gamma^1$ quadrupole:
\begin{eqnarray}
\label{P2RFB}
P_{\ell m}^{(2)} (\eta) &=& 
\sum_{i} p_{\ell,i\ell m} \\ \nonumber
&=&
- \frac{3}{4} \sqrt{\frac{(\ell+2)!}{(\ell-2)!}} 
\int_0^\eta d\eta' \, g(\eta',\eta) 
\int_0^{\eta'} d\eta'' \, \int_0^\infty dx \, x^2 \, 
\\ \nonumber
&\times& \sum_{i} \frac{2 R^{-3}}{j_{\ell+1}^2(k_{i\ell} R)}
S_{\ell m} (x,\eta',\eta'')  \, j_\ell (k_{i\ell} x)   \, 
\frac{j_\ell (k_{i\ell} \Delta \eta)}{(k_{i\ell} \Delta \eta)^2} \, 
j_2 (k_{i\ell} \Delta\eta')  \; .
\end{eqnarray}
Notice that now, making $r_1=x$, $r_2=\Delta\eta$ and
$r_3=\Delta\eta'$ we have on the right hand side
a spacetime window function given by:
\be
\label{WFB}
\tilde{W}^3_\ell (r_1,r_2;r_3) = \pi \, R^{-3}
\frac{r_1^2}{r_2^2} \sum_{i=1}^\infty \frac{1}{k_{i \ell}^2 }
\frac{j_\ell (k_{i \ell} \, r_1) \, j_\ell (k_{i \ell} \, r_2) \, j_2 (k_{i \ell}\,  r_3)}
{ j_{\ell+1}^2 (k_{i \ell}\,  R)} \; .
\ee
Although we haven't been able to prove mathematically that this
Fourier-Bessel window function is indentical to the Fourier window
function of Eq. (\ref{IK3}), we have checked numerically that they
are identical -- including the factor of $\pi$ which relates the phase
spaces of the two basis functions. Hence, the lowest-order
contribution to CMB polarization that results from using the Fourier-Bessel
representation is again given, precisely, by Eq. (\ref{P2R2}).

In fact, we can prove (see the Appendix) that the not only the
two window functions $W^3_\ell$ and $\tilde{W}^3_\ell$ are
equal, but that {\it all} the window functions derived in Section III are 
identical to the window functions that arise in the Fourier-Bessel 
expansion, if a certain generalization of the
orthogonality relation, Eq. (\ref{OBB}), is valid:
\be
\label{OBB2}
\sum_{i=1}^\infty \frac{j_{\ell'} (k_{i \ell} \, r_1) 
\, j_{\ell'} (k_{i \ell} \,  r_2)}{j_{\ell+1}^2 (k_{i \ell} \,  R)}
= \frac12 \, R^3 \, r_1^{-2} \, \delta(r_1-r_2) \; .
\ee
We have checked numerically that this relation seems to hold true
for a range of $\ell$ and $\ell'$, and for any arguments 
$0 \leq r_{1,2} < 1$, but as far as we know this has not been proven 
anywhere in the literature about Bessel
functions -- even though it clearly is a fundamental tool for relating 
quantities in the Fourier and in the Fourier-Bessel expansions.

We have also checked that the integral equations (\ref{TMB})-(\ref{PMB}) 
lead to the usual hierarchy of Boltzmann equations 
\cite{Ma:1995ey,Seljak:1996is,Hu:1997hp}, where now
there is one full independent hierarchy for each mode $\{ i \ell m \}$.
It is curious that, while in the
usual Fourier analysis what generates the hierarchy of Boltzmann 
equations are the recursion relations of the Legendre polynomials 
in the angular dependence $\hat{k}\cdot\hat{l}$, in our case the 
generators of the hierarchy are the recursion relations of the 
radial modes -- the spherical Bessel functions. Since
the two special functions are intimately related by Rayleigh's expansion
of the plane wave (which is ultimately what regulates the line-of-sight
integrals), it is indeed natural that both basis could be used to generate that
hierarchy.

The crucial difference between the Fourier-Bessel
series and the usual Fourier analysis lies in the discrete momenta $k_{i\ell}$
that can contribute to the observables $\Theta_{\ell m}$ and 
$P_{\ell m}$ in the Fourier-Bessel expansion. 
Critically, in this discretized series the first mode to contribute at each 
multipole is $k_{1\ell} \sim \ell/R$. 
Another feature is that the total number of modes
which one needs to compute to obtain anisotropies up to some $\ell_{max}$
is $N_{max} \sim \ell_{max}^2/9$ (for $\ell_{max} \gg 10$).

There is, however, an apparent drawback of the Fourier-Bessel basis: 
although it clearly is a superior method to solve our sort of initial values 
problem when compared to the Fourier basis, when the
underlying spatial fields are Gaussian the coefficients of the 
Fourier-Bessel series do not obey simple statistics like those of the 
Fourier modes. We will now turn to these issues.

\subsection{Statistics and power spectra in the Fourier-Bessel basis}

Take a Gaussian field $f(\vec{x})$ in $\mathbb{R}^3$, which
is expanded into spherical harmonics in position and in Fourier
space as in Eqs. (\ref{FLM}). If homogeneity and isotropy are
unbroken, the two-point correlation functions in position 
and in Fourier space can be expressed in terms of the radial
functions as:
\begin{eqnarray}
\label{CFR}
\langle f_{\ell m} (r) f^*_{\ell' m'} (r') \rangle &=& \delta_{\ell \ell'} \, 
\delta_{m m'}
\, \xi_\ell^f (r,r') \; ,
\\
\label{CFF}
\langle f_{\ell m} (k) f^*_{\ell' m'} (k') \rangle &=& 
\delta_{\ell \ell'} \, \delta_{m m'} \, k^{-2} \, P_f(k) \, \delta(k-k') \; .
\end{eqnarray}
Here the advantage of Fourier space becomes evident: 
translational invariance implies that the covariance matrix of the 
Fourier modes is completely diagonal.
The relationships between the position-space two-point correlation function
and the power spectrum are given by:
\be
\label{XIP}
\xi_\ell^f (r,r') = \frac{2}{\pi}
\int_0^\infty dk \, k^2 \, j_\ell (kr) \, j_\ell (kr') \, P_f(k) \; ,
\ee
and, conversely, by:
\be
\label{PIX}
P_f(k) = \int_0^\infty dr \, r^2 \, \frac{j_\ell(kr)}{j_\ell(kr')} \xi_\ell^f(r,r') \; .
\ee
From the identity above it is also evident that the two-point correlation 
function has a lot of redundant information, since many different
traces of it can lead to the power spectrum. If the field $f$
is Gaussian, these correlation functions are the only non-trivial
statistical momenta of that field's distribution function.

We want to obtain the corresponding relations for the modes of
the Fourier-Bessel series. This can be easily achieved 
by taking fields $f(\vec{x})$ in $\mathbb{R}^3$ and passing them
through a radial window function $W(r)$, such that $W(r\geq R) =0$ 
-- e.g., the tophat window function:
\be
\label{TWF}
W^{TH}(r) = \theta(R-r) \quad , \quad 
W^{TH}(k) = \sqrt{\frac{2}{\pi}} \, \frac{R^3}{kR} \, j_1(kR) \; ,
\ee
where $\theta(x)$ is the step (Heaviside) function.
In this way, the filtered functions will obey the boundary 
conditions, $\tilde{f}(R)= W(R) f(R) = 0$.

In Fourier space, the effect of a window function is to couple 
the different modes:
\be
\label{FWI}
\tilde{f} (\vec{k}) = \int\, \frac{d^3 q}{(2\pi)^{3/2}} 
\, W(\vec{k}-\vec{q}) \, f (\vec{q}) \; .
\ee
Hence, the Fourier transform of the filtered function 
acquires non-diagonal correlations $k$-space:
\be
\label{SPP}
\langle \tilde{f} (\vec{k}) \tilde{f}^* (\vec{k'}) \rangle =
\int \frac{d^3 q}{(2\pi)^3} \, W(\vec{k}-\vec{q}) \,W(\vec{k'}-\vec{q}) \, P_f(q) \; .
\ee

For a purely radial window function $W(r)$, we obtain with
the help of Eqs. (\ref{HTR}) that the spherical
harmonic components of the filtered function are:
\be
\label{SFF}
\tilde{f}_{\ell m} (k) = \int dq \, q^2 \, f_{\ell m} (q) \, W_\ell (k,q) \; ,
\ee
where $W_\ell(k,q)$ is the (symmetric) mode-coupling kernel of the 
radial window function:
\be
\label{SF2}
W_\ell (k,q) = \frac{2}{\pi} \, \int_0^\infty dr \, r^2 \, W(r) \, j_{\ell} (kr) \, j_{\ell} (qr) \; .
\ee
Hence, it is clear that if $W(r) \rightarrow 1$ then $W_\ell(k,q) 
\rightarrow q^{-2} \delta(k-q)$ and
we recover the Fourier modes of the $\mathbb{R}^3$ field. 
For a generic radial window function, however, there will be mixing of modes,
and the covariance matrix will be non-diagonal. 
The filtered spectrum is then related to the physical power spectrum
through:
\be
\label{FPS}
\langle \tilde{f}_{\ell m} (k) \tilde{f}^*_{\ell m} (k') \rangle
= \int_0^\infty dq \, q^2 \, W_\ell(k,q) \, W_\ell(k',q) \, P_f(q) \; .
\ee
In particular, for the Fourier-Bessel modes, which are
related to the (filtered) spherical modes in Fourier space by
$ f_{i\ell m} = \sqrt{2\pi} \, i^\ell \, j_{\ell+1}^{-2} (k_{i\ell} R) \tilde{f}_{\ell m} (k_{i\ell})$, this last identity implies that:
\begin{eqnarray}
\nonumber
\langle f_{i \ell m} f^*_{j \ell m}  \rangle
&=& \frac{2\pi \, R^{-6}}{j_{\ell+1}^2 (k_{i\ell} \, R) \, 
j_{\ell+1}^2 (k_{j\ell}\,  R)}
\int_0^\infty dq \, q^2 \, W_\ell (k_{i\ell},q) \, W_\ell (k_{j\ell},q) \, P_f(q) 
\\
\nonumber
&=&   \int_0^R dr \, r^2 \, 
\frac{2 R^{-3} \, j_\ell(k_{i\ell} \, r)}{j_{\ell+1}^2 (k_{i\ell}\,  R)}
\int_0^R dr' \, {r'}^2 \,
\frac{2 R^{-3} \, j_\ell(k_{j\ell}\,  r')}{j_{\ell+1}^2 (k_{j\ell} \, R)}
\, \times \frac{2}{\pi} \, \int_0^\infty dq \, q^2 \, P_f(q) \, j_\ell(q r) \, j_\ell(q r')
\\
\label{CFB}
&=&  \int_0^R dr \, r^2 \, 
\frac{2 R^{-3} \, j_\ell(k_{i\ell} \, r)}{j_{\ell+1}^2 (k_{i\ell} \, R)}
\int_0^R dr'  \, {r'}^2 \, 
\frac{2 R^{-3} \, j_\ell(k_{j\ell} \, r')}{j_{\ell+1}^2 (k_{j\ell} \, R)} \,
\xi_\ell^f (r,r') \; ,
\end{eqnarray}
where we have used the tophat window function from the first
to the second line. These expressions show how to compute the
covariance of the Fourier-Bessel modes from either the power spectrum
or from the two-point correlation function in position space.
It is also useful to obtain the equivalent of Eq. (\ref{XIP}) in the 
Fourier-Bessel representation. By the completeness relation, 
Eq. (\ref{OBB}), it is easy to see that:
\be
\label{FBPK}
\sum_{i} \sum_{j} \,
j_\ell(k_{i\ell} r) \, j_\ell(k_{j\ell} r')
\langle f_{i\ell m} f_{j\ell m}^* \rangle = \theta(R-r) \, \theta(R-r') \, \xi_\ell^f (r,r') \; ,
\ee
where $\theta(r)$ is the step (Heaviside) function.

The two-point correlation function (as opposed to the 
Fourier spectrum) is more directly related to the physical 
observables, since it remains
invariant as long as we keep within the causally accessible region 
(i.e., $r \leq R$ and $r' \leq R$.) 
However, Eq. (\ref{CFB}) also tells us that the two-point function of the 
Fourier-Bessel modes has non-diagonal terms  (albeit only in $k$-space.) 
This does not pose a problem, because the spectrum is not really
an observable: it can only be estimated (with exactly the same tools and
assumptions as usual) from observables such as the temperature
and polarization maps, as well as their derived products such as the 
angular power spectra. And since we saw in the previous section that the
observables retain exactly the same relations to the sources of 
anisotropies as they do in the usual Fourier expansion in $\mathbb{R}^3$,
we conclude that the Fourier-Bessel expansion fulfills all the
requirements to faithfully express the physics of the CMB.

\subsection{Angular power spectra}

Most of the useful cosmological information that we get from the 
CMB comes from the angular power spectra, because
of their simple relationship with the Fourier power spectrum.
For a function $f(\vec{x})$, the angular spectrum at radius $r$
can be defined from Eq. (\ref{CFR}), as $C_\ell^f (r) = \xi_\ell^f (r,r)$.
Below we show that in the Fourier-Bessel expansion the angular
power spectra assume exactly the same values as they would 
if we did not assume that the space was limited to the sphere $r \leq R$.

The argument is simplest for the temperature anisotropies
to order $\gamma^1$, and generalizes in a trivial manner to
the higher-order terms.
The angular power spectrum for temperature
in the Fourier-Bessel case reads:
\begin{eqnarray}
\label{TTF}
\langle \Theta_{\ell m}^{(1)} \Theta_{\ell' m'}^{(1)*} \rangle &=& 
\langle \sum_{i} \, \Theta_{i \ell m}^{(1)} 
\, \sum_{i'} \, \Theta_{i' \ell' m'}^{(1)*} \rangle 
=C^{TT \, (1)}_\ell (\eta) \, \delta_{\ell \ell'} \, \delta_{m m'} \; .
\end{eqnarray}
Using the orthogonality conditions of the Fourier-Bessel 
basis in target space,  Eq. (\ref{OBB}), we obtain:
\begin{eqnarray}
\label{CTB}
C_\ell^{TT \, (1)} (\eta) &=& 
\int^\eta d\eta_1' \int^\eta d\eta_2' \; \langle 
S_{\ell m} (x=\Delta\eta_1,\eta_1')
S^*_{\ell m} (x=\Delta\eta_2,\eta_2')
\rangle 
\\ 
\nonumber
&=& 
\langle 
\bar{S}_{\ell m}
\bar{S}^{*}_{\ell m}
\rangle_{PLC}
\; .
\end{eqnarray}
But this is exactly the usual result: to lowest order, the temperature
angular power spectrum is given by the average over the PLC
of the (angular) two-point angular correlation function of the Sachs-Wolfe, 
Doppler and integrated Sachs-Wolfe source terms, properly
weighted by the visibility.
For the $\gamma^2$ and higher-order terms, the procedure is
precisely the same and leads back to the same relation
between the angular power spectra and the sources as
happened in position space, where the spacetime window functions
regulate which sources contribute to the anisotropies in the space
and time integrals. Hence, we have shown that not only the observables
(the temperature anisotropies), but also that the statistics of the
angular power spectra in the Fourier-Bessel expansion are identical 
to the usual case of the Fourier expansion.

\subsection{Fourier {\it v.} Fourier-Bessel}

The identity between the spacetime window functions at all orders 
in $\gamma$ implies that the source and the observables 
(the temperature and polarization maps, or equivalently their 
spherical harmonic components $\Theta_{\ell m}$ and 
$P_{\ell m}$) are related in exactly the same way in the 
Fourier-Bessel basis and in the Fourier basis. The statistics of 
the angular power spectra, therefore, are also related to
the statistics of the underlying matter and metric fields in precisely 
the same way in the two representations.

Hence, the Fourier-Bessel basis is, in some respects, completely 
equivalent to the Fourier basis: it represents the same physics and
it expresses the same observables as its Fourier counterparts -- 
it even has the same statistics. However, in at least one respect
the Fourier-Bessel basis is superior to the Fourier basis: it has a precise
prescription for the discretized tower of modes that contribute 
for the observables at each multipole.
These modes take into account exactly the relevant pieces of information
from the sources, the ones that propagate from the initial value surface 
to the physical observables -- no more, no less. And the statistics 
of the power spectra, as we have demonstrated above, is related 
in precisely the same way to the statistics of the (presumably 
Gaussian) matter fields, just as happens in the usual analysis in 
Fourier space. 

Finally, as an initial-value formulation the Fourier-Bessel expansion is
vastly superior to the Fourier representation because it does not
waste any resources keeping track of irrelevant variables such as 
super-Hubble modes or modes which trace out of the observable.
All the information is encoded in a discrete series of momenta, and
we do not have to guess how to subdivide the Fourier space in
sufficiently small pieces in order to sample the observables we 
want to compute -- the Fourier-Bessel modes already provide the 
unique, optimal choice.

\section{Conclusions}

In this paper we have shown how causality constraints in position space
regulate which sources of anisotropies (the matter and metric perturbations) 
can contribute to the CMB.
This causal structure is manifested order by order in a series of terms 
corresponding to the number of interactions that photons experienced 
over the past light-cone of the observer -- or, equivalently, a power
series on the visibility $\gamma = e^{-\mu}$.

When expressed in position space, the line-of-sight integrals 
acquire an intuitive interpretation in terms of scatterings over the 
light-cones of the successive scatterings, all the way from the sources to the
point and time of observation. 
In particular, we find that, in position space, only the sources of anisotropy 
that are inside our past light-cone are taken into account. This statement
is exact to all orders -- as it should be, since the causal nature of the propagation of
photons is the key ingredient in the line-of-sight integrals from which we
started.

At each order in the power series on the visibility, the sources are weighted
by spacetime window functions. These window functions can be
complicated for a high number of scatterings, but they all obey
a very simple rule: they vanish indentically 
unless some extremely simple sets of inequalities are satisfied. These
inequalities have a simple geometrical interpretation: if the position
of the source and the radii of the light-cones of the interactions
cannot form a flat polygon, the spacetime window functions vanish.
An interesting question which we did not have time to address is 
at what number of scatterings prior to free streaming we can cut off this
series so that the error in the temperature distribution is, say, of
order 1\%.

One of the implications of these causality constraints is that, 
whatever the properties of the Universe outside a
limiting radius $R$, the source fields do not propagate to the CMB
observables -- and the line-of-sight integrals both in Fourier space and
in the Fourier-Bessel expansion retain this property.
In practice, this means that we can use the Fourier-Bessel framework
to compute the CMB -- even though it puts the Universe in a ``spherical box'',
and discards all the information outside of that box.

In the Fourier-Bessel basis, the fields
are decomposed in spherical harmonics and a series of 
discrete eingenmodes $k_{i\ell}$ (as opposed to the continous
modes of Fourier space.)
The first eigenmode for each multipole $\ell$ is $k_{1\ell} \sim \ell/R$.
CMB observables are exactly the same as in the Fourier
basis -- but the Fourier-Bessel basis is optimal, in the sense
that it does not keep track of irrelevant modes, only the ones
that contribute constructively to the physical observables.

The previous discussion implies that our results and methods are 
suitable for analytical and numerical
studies of CMB temperature and polarization maps  
in models with large-scale inhomogeneities, statistical anisotropy or 
non-gaussianities of any kind 
kind \cite{Liguori:2003mb,Komatsu:2003iq,Liguori:2007sj,Elsner:2009md}. 
Our results may be useful also in simulations of the CMB in the 
presence of topological defects. It is not clear whether the
methods described here can be employed to study models with 
non-trivial topology \cite{Cornish:2003db}, since in those cases 
the Fourier or Fourier-Bessel basis functions may not be eigenvectors 
of the Laplace-Beltrami operator.

%%%%%%%%%%%%%%%%%%%%%%%%%%%%%%%%%%%%
\subsection*{Acknowledgements}
%%%%%%%%%%%%%%%%%%%%%%%%%%%%%%%%%%%%

The authors would like to thank J. C. A. Barata for many conversations
on the properties of Bessel functions, and to Mathias Zaldarriaga for 
useful comments.
This work was supported by FAPESP and CNPq.

\bibliographystyle{h-physrev}
\bibliography{paper_cmb_los}

%%%%%%%%%%%%%%%%%%%%%%%%%%%%%%%%%%%%%%%%%%%%%%

\appendix

\section{Integrals of Bessel Functions}
\label{ApInts}

%%%%%%%%%%%%%%%%%%%%%%%%%%%%%%%%%%%%%%%%
%%%%%%%%%%%%%%%%%%%%%%%%%%%%%%%%%%%%%%%%

\subsection{Integral of products of spherical Bessel functions}

We will now compute the integrals of spherical Bessel functions
that were presented in Section \ref{SecCau} to obtain
the spacetime window functions. Parts of the methods
used here can be found in \cite{Watson}.
The results below also provide the motivation for 
the diagrammatic representation of the
window functions shown in Fig. \ref{Polygons}.

The spherical Bessel functions are associated with the matrix elements
of the Euclidean group in three dimensions, $E_3$ \cite{Talman}. The
Euclidean group $E_3$ consists of the set of transformations that leaves
spatial distances invariant -- i.e., spatial translations and rotations.
The rules of group multiplication lead to addition theorems for the
special functions which realize the group representation, one example of
which is the orthogonality condition of Eq. (\ref{OBF}). 
Although $E_3$ is not compact, the spherical Bessel functions also 
obey an addition rule, namely \cite{Talman}:
\begin{equation}
\label{bessel_addition}
\frac{j_m(kr)}{(kr)^m} = 
\sum_{n=m}^{\infty} \frac{(2n+1)\,  j_n(kr_1) \, j_n(kr_2)}{[(kr_1)(kr_2) \sin \theta]^m}
\, P_n^(m)(\cos \theta) \, ,
\end{equation}
where $m$ is even.

\subsection{Integral of three spherical Bessel functions}
\label{drei}

We can use Eq. (\ref{bessel_addition}) 
and the orthogonality of Legendre polynomials,
\begin{equation}
\int_{-1}^{1}  dx \, P_\ell^{(m)}(x)\,  P_{\ell'}^{(-m)} (x) = \frac{2}{2\ell+1} \delta_{\ell, \ell'} \, ,
\end{equation}
to reduce the product of two Bessel functions to only one Bessel
function. Choosing $m=2$ due to the demands of our particular problem,
we have:
\begin{eqnarray}
\label{parte_1}
j_{\ell} (k r_1)\,  j_\ell (k r_2) & = & 
\sum_{\ell'=2}^{\infty} j_{\ell'} (k r_1) \, j_{\ell'} (k r_2) \, \delta_{\ell, \ell'} \nonumber\\
& =& \sum_{\ell'=2}^{\infty} \frac{2\ell'+1}{2} \int_{-1}^1 d(\cos \alpha) \, 
j_{\ell'} (k r_1) \, j_{\ell'} (k r_2)
P_{\ell}^{(2)} (\cos \alpha) \, P_{\ell'}^{(-2)} (\cos \alpha) \nonumber\\
& = &\frac{1}{2} \int_{-1}^1 d(\cos \alpha) \, P_{\ell}^{(-2)} (\cos \alpha) 
\, (kr_1)^2 \, (kr_2)^2 \, \sin^2 \alpha 
\nonumber \\ 
& \times &
\sum_{\ell'=2}^{\infty} \frac{(2\ell'+1)\,  j_{\ell'} (k r_1)\,  j_{\ell'} (k r_2) 
\, P_{\ell'}^{(2)} (\cos \alpha)}{[(kr_1)(kr_2) \sin \alpha]^2} \; .
\end{eqnarray}
Now we perform a change of variable, calling $r$ the side of the triangle
whose other two sides are $r_1$ and $r_2$, such that $\alpha$ is the 
angle between $r_1$ and $r_2$, i. e. 
$r^2 = r_1^2 + r_2^2 - 2 r_1 r_2 \cos \alpha$. Performing this change of
variables, we obtain:
\begin{displaymath}
\int_{-1}^{1} d(\cos \alpha) \to \int_{|r_1-r_2|}^{r_1+r_2} d r \frac{r}{r_1 r_2} \, ,
\end{displaymath}
and therefore:
\begin{equation}
\label{parte_2}
j_{\ell} (k r_1) \, j_\ell (k r_2) = \frac{1}{2} \int d r \,  k^2 \, \frac{r_1 r_2}{r} 
\, j_2(kr) \, P_{\ell}^{(-2)} (\cos \alpha) \, \sin^2 \alpha \; .
\end{equation}

Consider, then, the integral that is relevant to us:
\begin{equation}
\int dk \, j_\ell(kr_1)
\, j_\ell(k r_2) \, j_2(kr_3) = \frac{r_1 r_2}{2} \int \, \frac{dr}{r} 
P_{\ell}^{(-2)} (\cos \alpha) \, \sin^2 \alpha
\int dk \, k^2 \, j_2(kr) \, j_2(kr_3)\; .
\end{equation}
But now we can employ the orthogonality of Bessel functions, Eq. (\ref{OBF}), 
so that the $k$ integral gives $(\pi/2) r^{-2} \delta(r-r_3)$ and
the radial integral can be computed to arrive at the final expresion:
\begin{equation}
\label{tres_bessels}
I^{(3)}_\ell (r_1,r_2,r_3)
= \int dk \, j_{\ell} (k r_1) \, j_\ell (k r_2) \,  j_2 (k r_3) = \frac{\pi}{4} \frac{r_1 r_2}{r_3^3}
P_{\ell}^{(-2)} (\cos \alpha) \, \sin^2 \alpha \; ,
\end{equation}
where $r_1$, $r_2$ and $r_3$ must form a triangle: if they do not,
the radial integral yields zero because then $r_3$ cannot be equal to 
some $r$ which, by assumption, forms a triangle together with $r_1$ and
$r_2$. Identifying $r_1 \rightarrow x$, $r_2 \rightarrow
\Delta\eta$ and $r_3 \rightarrow \Delta \eta'$
we obtain the result shown in Eq. (\ref{IK3}).
This integral is also computed in \cite{Watson}, in 
a more general case but employing other methods.

For the series representation of the window function, consider
Eq. (\ref{WFB}). The same trick that was shown above, i.e., to 
exchange two spherical Bessel functions for an integral over
a Legendre polynomial, can be used to obtain:
\be
\label{3BFS}
\sum_{i}  
\frac{ j_\ell (k_{i\ell} r_1) \,j_\ell (k_{i\ell}  r_2) \,j_2 (k_{i\ell}  r_3)}
{k_{i\ell}^2 \, j_{\ell+1}^2 (k_{i\ell} R) }
= \,
\frac{r_1 r_2}{2} \, \int \frac{dr}{r} P_{\ell}^{(-2)} (\cos\alpha) \sin^2\alpha
\times \frac{1}{r} \sum_{i} \frac{j_2 (k_{i\ell} r_3) j_2 (k_{i\ell} r)}
{j_{\ell+1}^2 (k_{i\ell} R)} \; .
\ee
Using now the conjectured orthogonality relation, Eq. (\ref{OBB2}), with
$\ell'=2$, we obtain that the window function of Eq. (\ref{WFB}) is
indeed identical to the window function of Eq. (\ref{IK3}).

\subsection{Integration of four spherical Bessel functions}

Consider now the integral that appears in the case of $N=2$ scatterings:
\begin{equation}
\label{vier}
I^{(4)}_\ell = \int_0^{\infty} dk \, k^{-2}  \, j_\ell(kr_1) \, 
j_\ell(kr_2) \, j_2(kr_3) \, j_2(kr_4) \, .
\end{equation}

To benefit from the results obtained above for the case of
the integral of three Bessel functions, let $\alpha_{12}$ be the
angle formed by $r_1$ and $r_2$, i.e., $r^2 = r_1^2 + r_2^2 - 2 r_1
r_2 \cos \alpha_{12}$, and use Eq. \eqref{bessel_addition} to rewrite
Eq. \eqref{vier} as:
\begin{eqnarray}
I^{(4)}_\ell &=&  \frac{r_1 r_2}{2}
\int_{|r_1-r_2|}^{r_1+r_2} \, 
\frac{dr}{r} \,  P_{\ell}^{-2} (\cos \alpha_{12}) \, \sin^2 \alpha_{12}  \,
\int_0^{\infty} dk  \, j_2(k r_3) \, j_2(kr_4) \, j_2(kr) \, 
\\ \nonumber
&=&  \frac{r_1 r_2}{2}
\int_{|r_1-r_2|}^{r_1+r_2} \, 
\frac{dr}{r} \,  P_{\ell}^{-2} (\cos \alpha_{12}) \, \sin^2 \alpha_{12}  \,
\times \, I^{(3)}_2 (r_3,r_4,r) \, 
\end{eqnarray}
The integral $I^{(3)}_2$ of three spherical Bessel functions of
order two is a symmetric case of Eq. \eqref{tres_bessels}, and it
vanishes unless $r$, $r_3$ and $r_4$ are the sides of a triangle. 
In this case we can choose any
angle in the triangle, so to be consistent let's choose that to be the angle 
between $r_3$ and $r_4$, that is,
$r^2 = r_3^2 + r_4^2 - 2 r_3 r_4 \cos \alpha_{34}$. 
Since $r_1$, $r_2$ and $r$ must also form a triangle,
$r$ and the angle $\alpha_{34}$ will be determined in terms of $r_1$, $r_2$,
$r_3$, $r_4$ and $\alpha_{12}$. The final answer is, therefore, given by:
\begin{equation}
\label{quatro_bessels_prev}
I^{(4)}_\ell (r_1,r_2,r_3,r_4) = \frac{\pi}{8} \, r_1 r_2 r_3 r_4 \,  \int_G \, 
\frac{dr}{r^4} \, P_{\ell}^{(-2)} (\cos \alpha_{12}) \, \sin^2\alpha_{12} 
\, P_{2}^{(-2)} (\cos \alpha_{34}) \, \sin^2 \alpha_{34} \, 
\end{equation}
where $G$ is the range of
values allowed for $r$ under the conditions that both the triangle of
sides $(r_1,r_2,r)$ and the one with sides $(r_3,r_4,r)$ exist. 
These conditions are
simply the set of inequalities that guarantee that the polygon of sides
$(r_1,r_2,r_3,r_4)$ can exist, i.e., $r_1 \leq r_2 + r_3 + r_4$ and the 
three other cyclical permutations of that inequality. 
The two triangles that must be formed so that 
Eq. \eqref{vier} does not vanish are shown in Fig. \ref{Polygons}.
This result was used in our Eq. (\ref{IK4}).

\subsection{The general case}

Our problem deals with the propagation of signals between points (events) 
in spacetime. The first signal propagates freely from the source to the
point where the photon first scatters, and then there is a set of
propagations from one scattering to the next, until finally there is a 
propagation term from the point where the photon have last scattered
to the point where it is observed. The propagation from the source to
the first scattering correspond to a term $j_2(k\Delta\eta_1)$ and each
propagation between scatterings to $j_2(k\Delta\eta_i)/(k\Delta\eta_i)^2$. 
The propagation from the last scattering to the
observation point corresponds to a term
$j_\ell(k\Delta\eta_N)/(k\Delta\eta_N)^2$. 
Besides these propagation terms, the Hankel
transforms (which were used to go back from Fourier to position space)
introduce a $j_\ell(kx)$ into our integral. 

Therefore, for $N$
scatterings we will get integrals over $k$ with an integrand 
having the following features:
\begin{quote}

$\bullet$ two Bessel functions of order $\ell$, 
of arguments $kr_1$ and $kr_2$ (by convention);

$\bullet$ $N$ spherical Bessel functions of order 2, of arguments
$k r_3, \ldots, k r_{N+2}$;

$\bullet$ a factor of $k^{-2}k^{-2(N-2)}=k^{-2(N-1)}$

\end{quote}

The method that was used above to compute the integrals 
in the cases $N=1$ and $N=2$
takes advantage of the fact that we can exchange pairs of spherical Bessel 
functions for Legendre polynomials and radial (or angular) integrals.
Now, we can do this for every pair of Bessel functions in the $N$-scattering
integral: if that number is even, every Bessel function can be exchanged
for an integral over a Legendre polynomial; if that number is odd, 
an extra Bessel function will appear. The final integral over $k$ can then be
computed with the help of the lower order integrals.

For $N=3$ this procedure leads to:
\begin{eqnarray}
\label{I5}
I^{(5)}_\ell &=&
\int dk \, k^{-4} \,
j_\ell(kr_1) \,
j_\ell(kr_2) \,
j_2(kr_3) \,
j_2(kr_4) \,
j_2 (k r_5)
\\ \nonumber
&=&
\frac{r_1 r_2 r_3 r_4}{2^2} 
\int \frac{dr_{12}}{r_{12}}
\, \int \frac{dr_{34}}{r_{34}} 
P_{\ell}^{(-2)} (\cos \alpha_{12}) \, \sin^2 \alpha_{12} \,
P_{2}^{(-2)} (\cos \alpha_{34}) \, \sin^2 \alpha_{34} \,
I^{(3)}_2 (r_{12},r_{34},r_5) \; ,
\end{eqnarray}
where $r_{34}$ makes a triangle together with $r_3$ and $r_4$, and
the angles are clearly indicated with respect to their respective sides.
Notice that, as opposed to the case $N=2$, when the sides and the
angle $\alpha_{12}$ uniquely determines the remaining angle of
that four-side polygon, in the case $N=3$ the triangle of sides 
$(r_{3},r_{4},r_{34})$ is totally free to acquire many shapes -- see also 
Fig. \ref{Polygons}. It is only when both $\alpha_{12}$ and $\alpha_{34}$
are given that the angle between $r_{12}$ and $r_{34}$ is fixed.

It is also instructive to look at the case $N=4$, for the integral
of six Bessel functions. In that case we have:
\begin{eqnarray}
\label{I6}
I^{(6)}_\ell &=&
\int dk \, k^{-6} \,
j_\ell(kr_1) \,
j_\ell(kr_2) \,
j_2(kr_3) \,
j_2(kr_4) \,
j_2 (k r_5) \,
j_2 (k r_6) \\ \nonumber
&=& 
\frac{r_1 r_2 r_3 r_4 r_5 r_6}{2^3} 
\int \frac{dr_{12}}{r_{12}}
\, \int \frac{dr_{34}}{r_{34}} 
\, \int \frac{dr_{56}}{r_{56}} 
\\ \nonumber
&\times&
P_{\ell}^{(-2)} (\cos \alpha_{12}) \, \sin^2 \alpha_{12} \,
P_{2}^{(-2)} (\cos \alpha_{34}) \, \sin^2 \alpha_{34} \,
P_{2}^{(-2)} (\cos \alpha_{56}) \, \sin^2 \alpha_{56} \,
I^{(3)}_2 (r_{12},r_{34},r_{56}) \; .
\end{eqnarray}
The expressions for $I^{(7)}_\ell$ and $I^{(8)}_\ell$ can be obtained
in terms of $I^{(4)}_\ell$; and so on.
With these methods it is trivial to compute the spacetime
window functions for an arbitrary number of scatterings.

The set of conditions under which the integrals above
are different from zero are those that ensure that each 
internal triangle (corresponding to each instance where two 
Bessel functions were exchanged for a Legendre polynomial 
and a Bessel function) exist. So, for $\ell=4$ we would impose:
\begin{eqnarray}
\nonumber
r_1 \leq r_2 + r_{12} \quad &,& \quad
r_2 \leq r_{12} + r_1 \quad , \quad
r_{12} \leq r_1 + r_2 \; , 
\\ \nonumber
r_3 \leq r_4 + r_{34} \quad &,& \quad
r_4 \leq r_{34} + r_3 \quad , \quad
r_{34} \leq r_3 + r_4 \; . 
\end{eqnarray}
With the additional condition that $r_{12}=r_{34}$ in the case
$\ell=4$ (see the discussion in A.3), 
it is trivial to verify that these conditions reduce to the 
inequalities (\ref{IN4}).

The set of conditions above simply tells us that the
four-sided polygon of Fig. (\ref{Polygons}) exists -- in
other words, that one can form a closed polygon with those sides.
For any $\ell$ the resulting set of conditions ensure that a 
flat polygon with the sides given by $r_1, \ldots, r_N$ exists, i.e.:
$$
r_1 \leq r_2 + \; \ldots \; + r_N \; ,
$$
and all cyclical permutations.
These inequalities constitute a simple set of constraints
that, if not satisfied, imply that the spacetime window functions
$W^N_\ell$ vanish identically. This simplifies tremendously the
integration of the sources over time in position space.
In particular, one of these inequalities imply that:
$$
x \leq \Delta\eta_1 + \; \ldots \; + \Delta\eta_{N-1} = \eta - \eta_N\; ,
$$
which means that all the sources that contribute to the
observables are located at radii $x$ which are {\it inside} 
the past light-cone of the
observation point at time $\eta$, all the way to the
time $\eta_N$ when those sources were evaluated, 
$N$ scatterings prior to the observation.

%%%%%%%%%%%%%%%%%%%%%%%%%%%%%%%%%%%%%%%%
%%%%%%%%%%%%%%%%%%%%%%%%%%%%%%%%%%%%%%%%

\subsection{Spacetime window function $W^3_\ell$}

The spacetime window functions regulate how sources at some
position $x$ contribute to the observables at time $\eta$.
For the case of one scattering (order $\gamma$), the window
function is a $\delta$-function on the PLC, $\delta(x-\Delta\eta)$,
where $\Delta\eta=\eta-\eta'$ and $\eta'$ is the time of the
scattering.

\begin{figure}[ht]
%\begin{centering}
\includegraphics[scale=0.65]{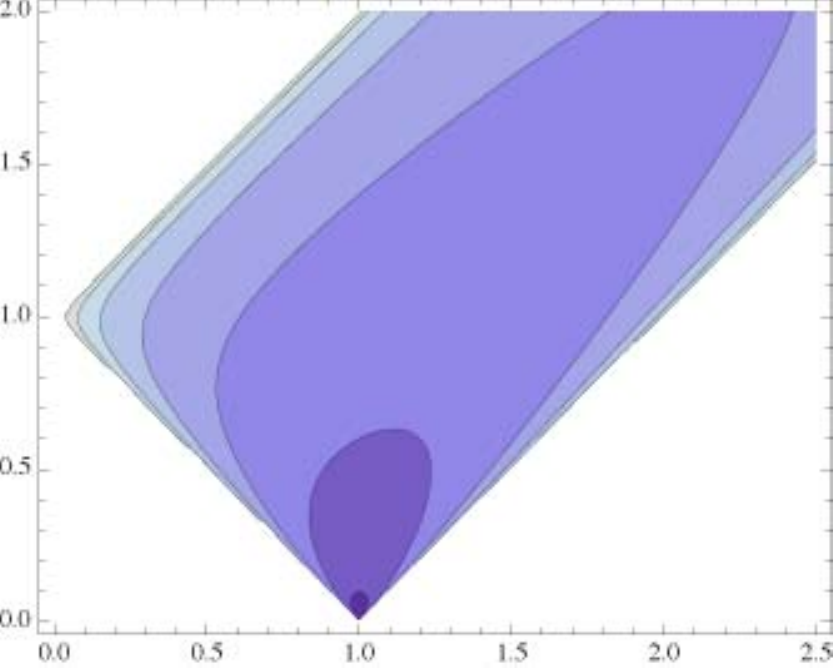}
\includegraphics[scale=0.65]{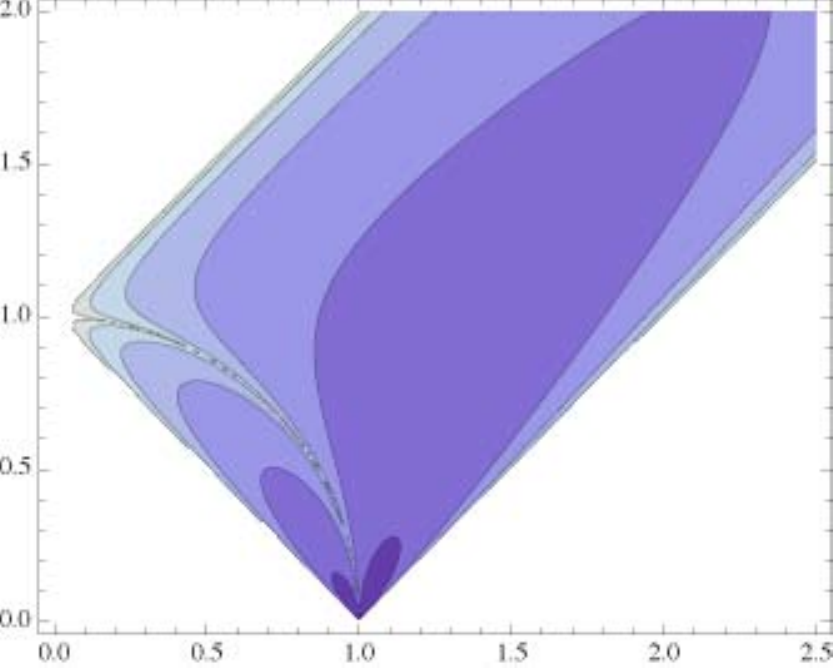}
\includegraphics[scale=0.65]{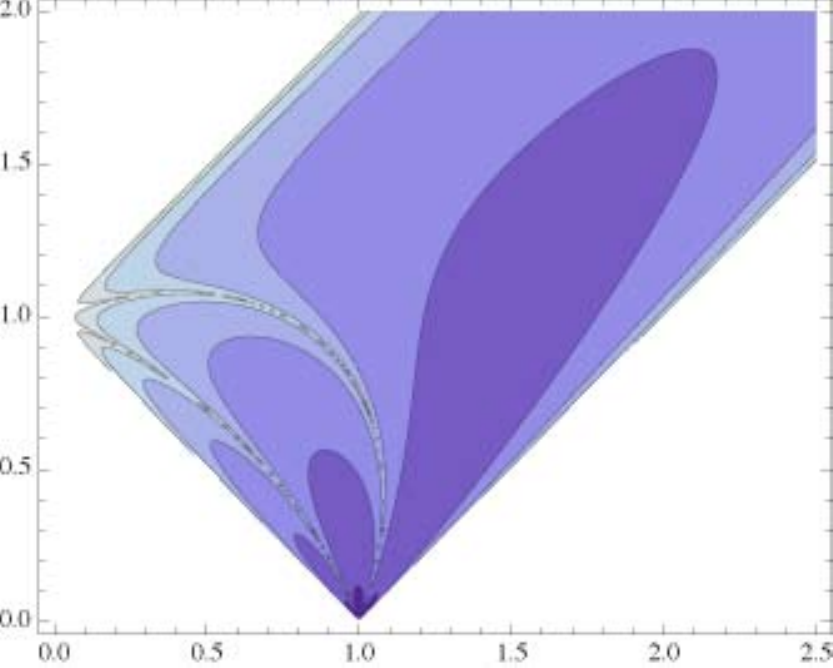}
\includegraphics[scale=0.65]{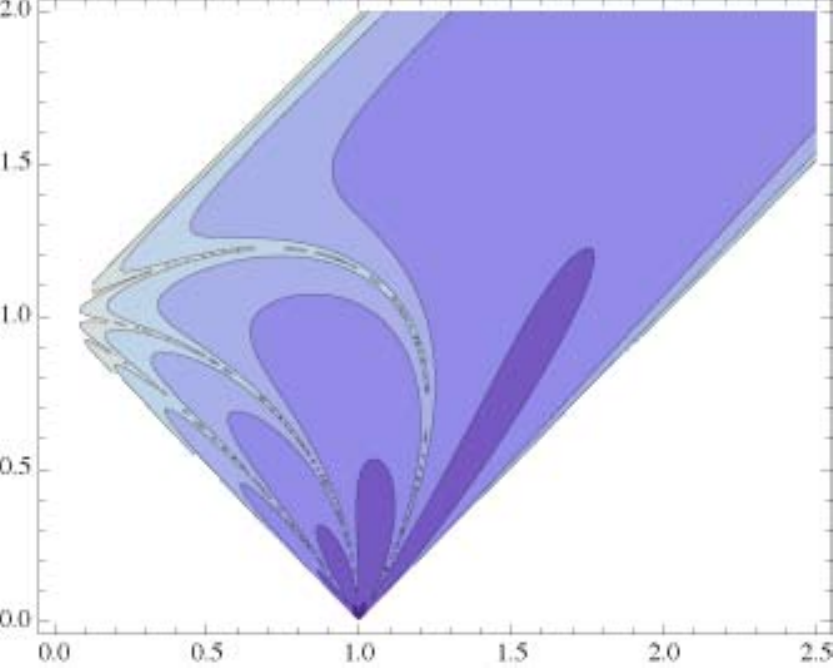}
\includegraphics[scale=0.65]{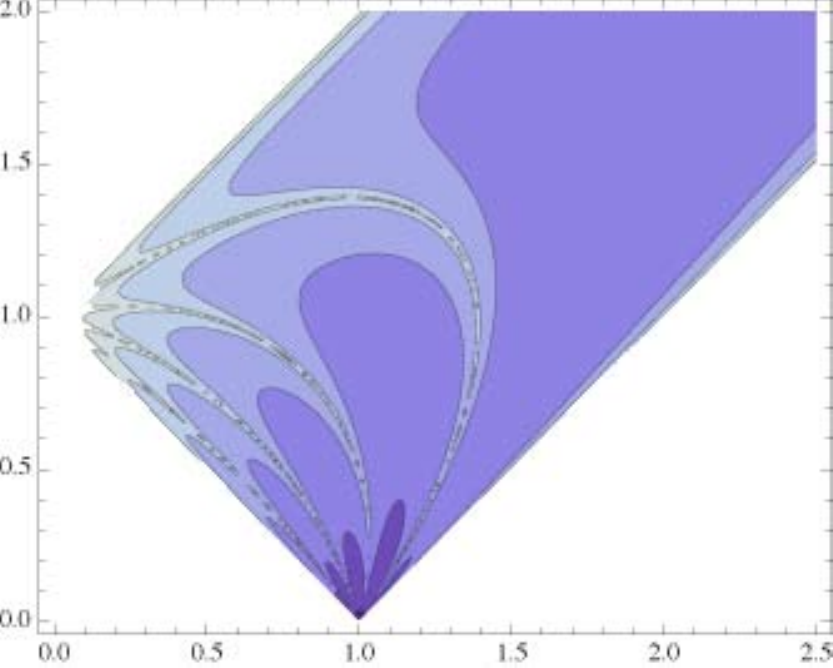}
\includegraphics[scale=0.65]{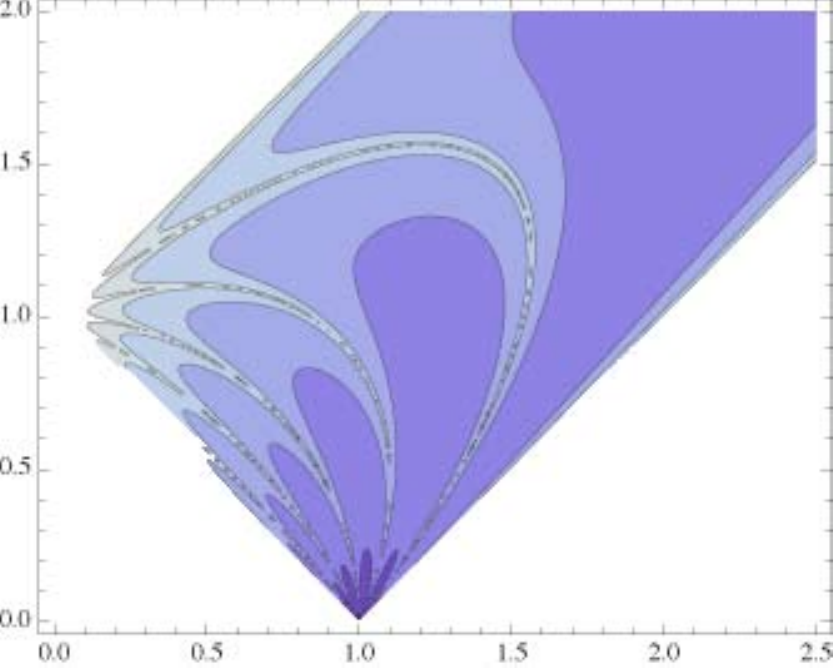}
%\par\end{centering}
\caption{ Contour plots of the
window functions $W^3_\ell(x,\Delta\eta;1)$ for the cases $\ell=$2, 3, 4 
(top panels, left to right panels), and 5, 6 and 7 (bottom panels.) 
In these plots $x$ (the radial position of the sources) corresponds to 
the horizontal axes, and $\Delta\eta = \eta-\eta'$ corresponds to the 
vertical axes. For visualization purposes we have fixed $\Delta\eta'=\eta'-\eta''=1$. 
Physically, this corresponds to taking sources at positions $x$ and times
$\eta''$, and photons which scatter at times $\eta'$ before they are observed
at time $\eta$.
For visualization purposes we plotted $\log|W^3_\ell(x,\Delta\eta;1)|$, so
large absolute values of the window functions are indicated by darker 
hues, and the window functions vanish in the white areas. Each lobe 
corresponds to intercalating negative and positive values of the window 
function.}
\label{W3c}
\end{figure}

\begin{figure}[ht]
%\begin{centering}
\includegraphics[scale=0.65]{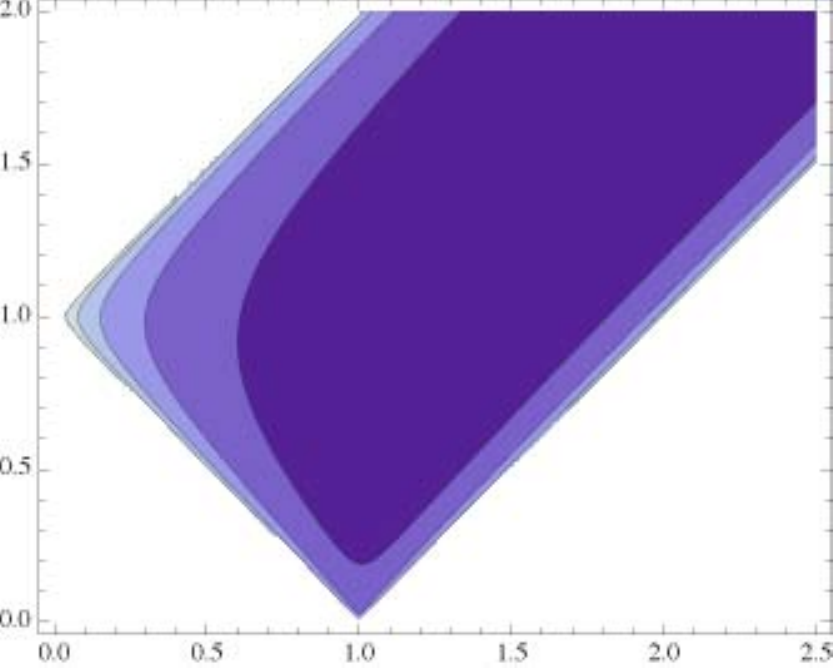}
\includegraphics[scale=0.65]{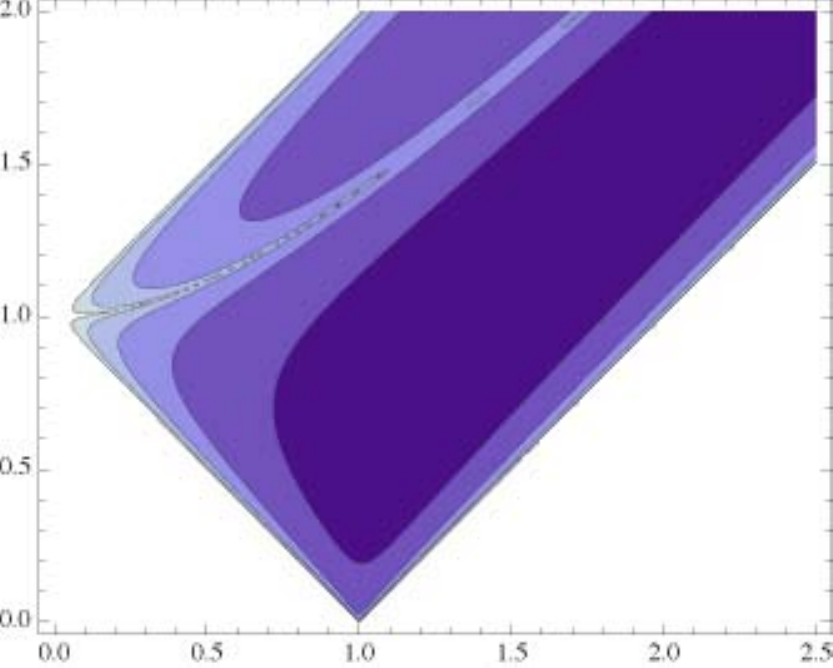}
\includegraphics[scale=0.65]{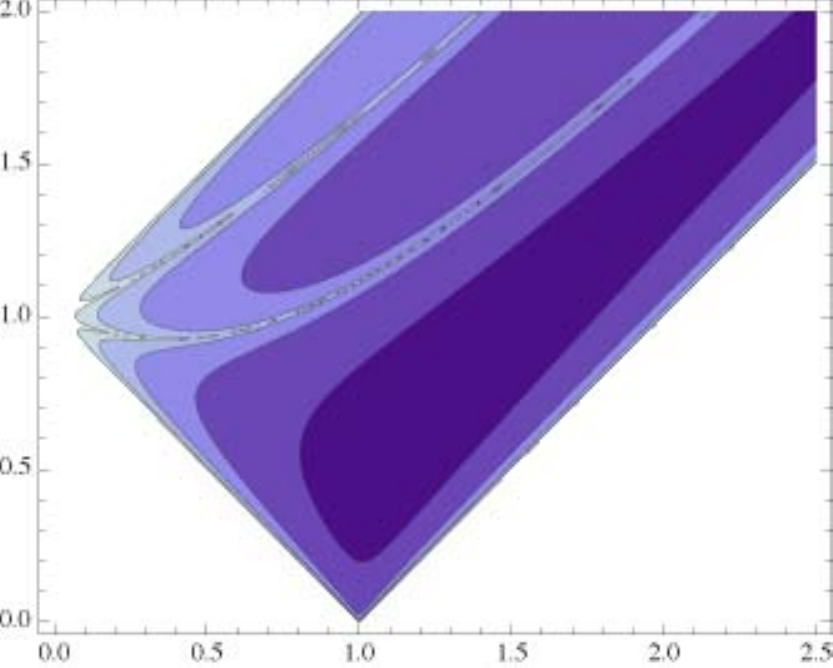}
\includegraphics[scale=0.65]{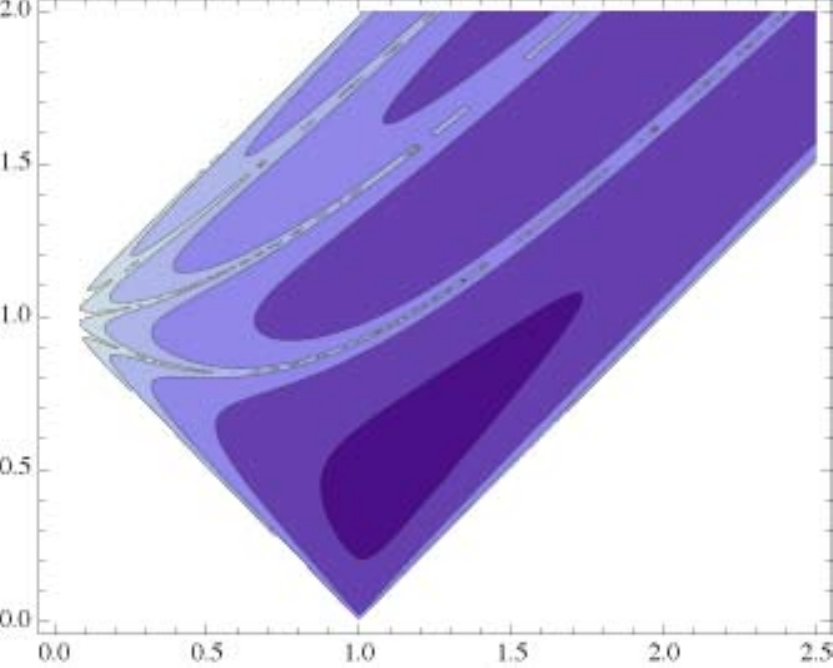}
\includegraphics[scale=0.65]{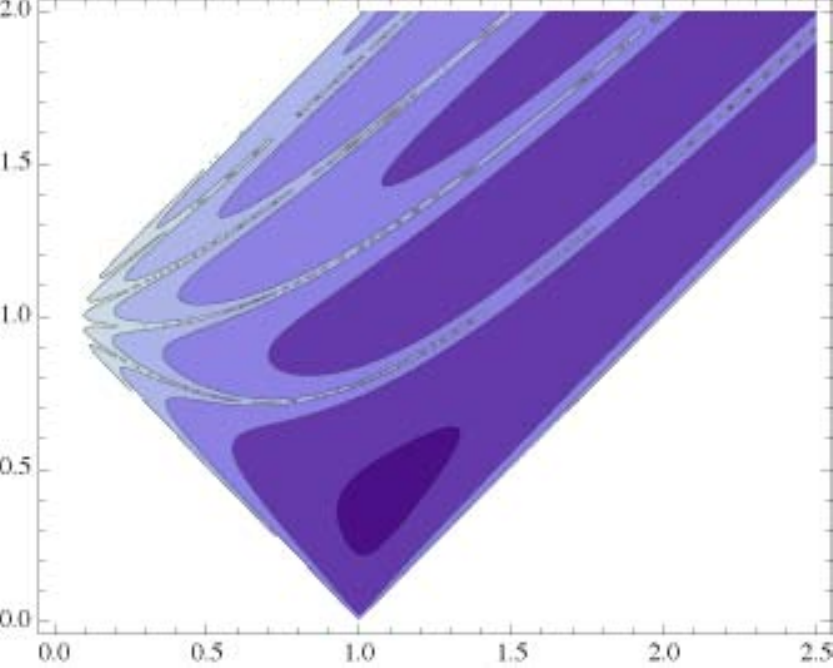}
\includegraphics[scale=0.65]{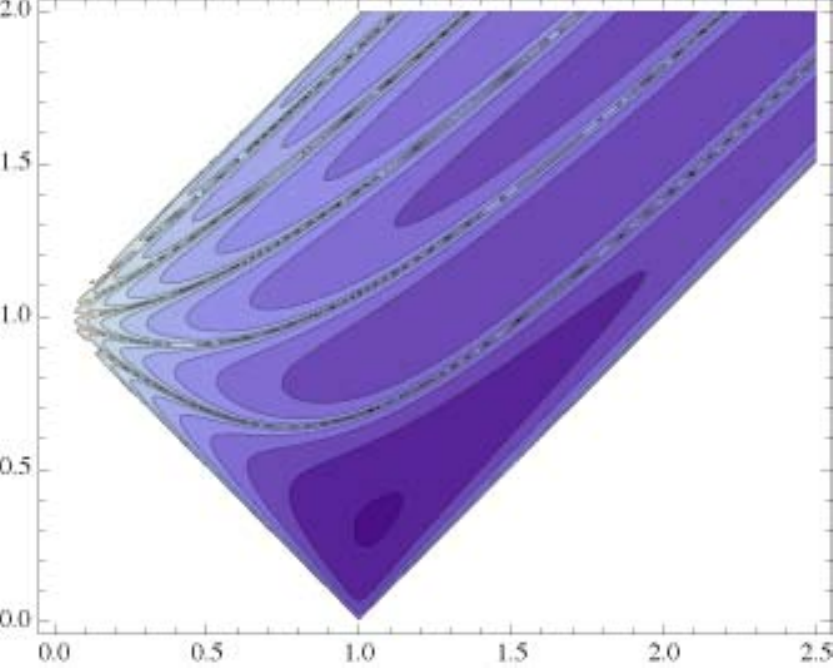}
%\par\end{centering}
\caption{ Same as above, but now we fix 
$\Delta\eta = \eta-\eta'=1$ so the contour plots correspond to
$W^3_\ell(x,1;\Delta\eta')$ for the cases $\ell=$2, 3, 4 
(top panels, left to right panels), and 5, 6 and 7 (bottom panels.) 
Now $\Delta\eta'$ corresponds to the vertical axes,
and $x$ corresponds to the horizontal axes.
}
\label{W3b}
\end{figure}

\begin{figure}[ht]
%\begin{centering}
\includegraphics[scale=0.65]{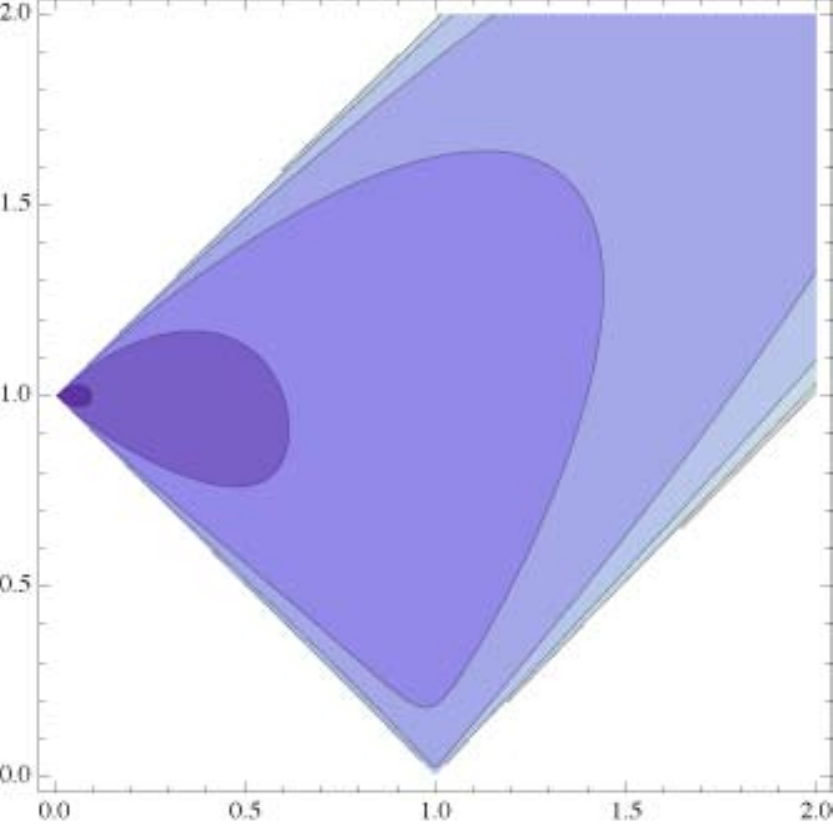}
\includegraphics[scale=0.65]{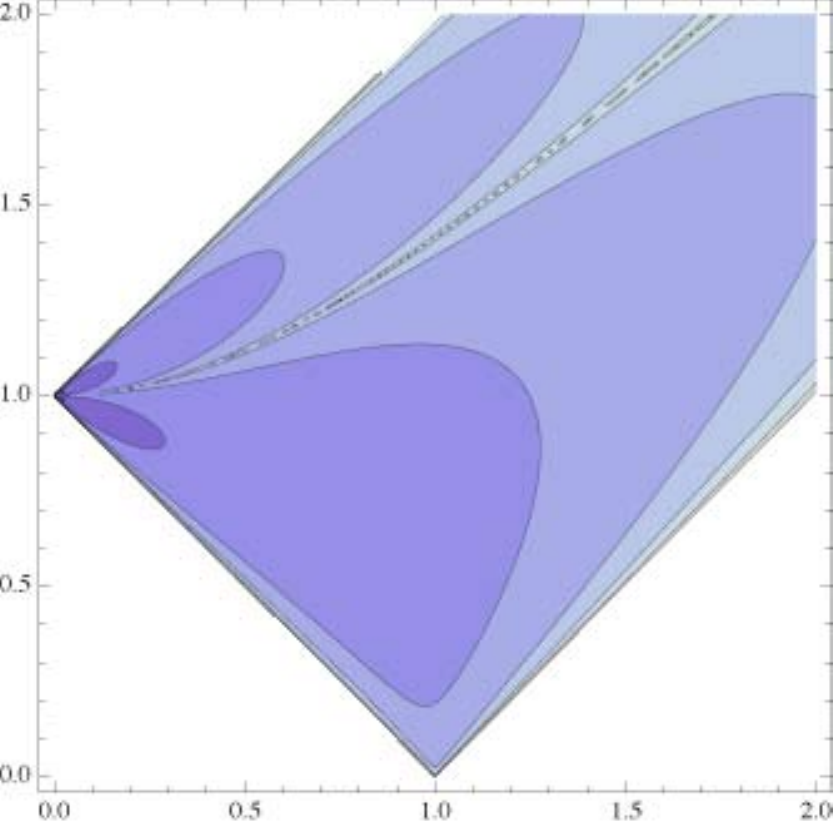}
\includegraphics[scale=0.65]{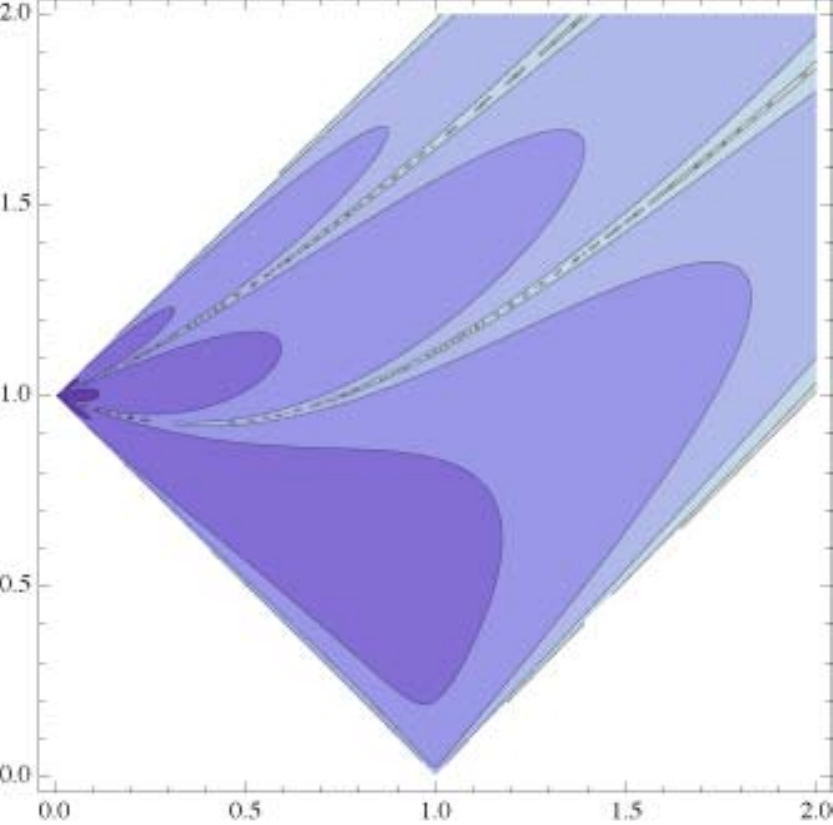}
\includegraphics[scale=0.65]{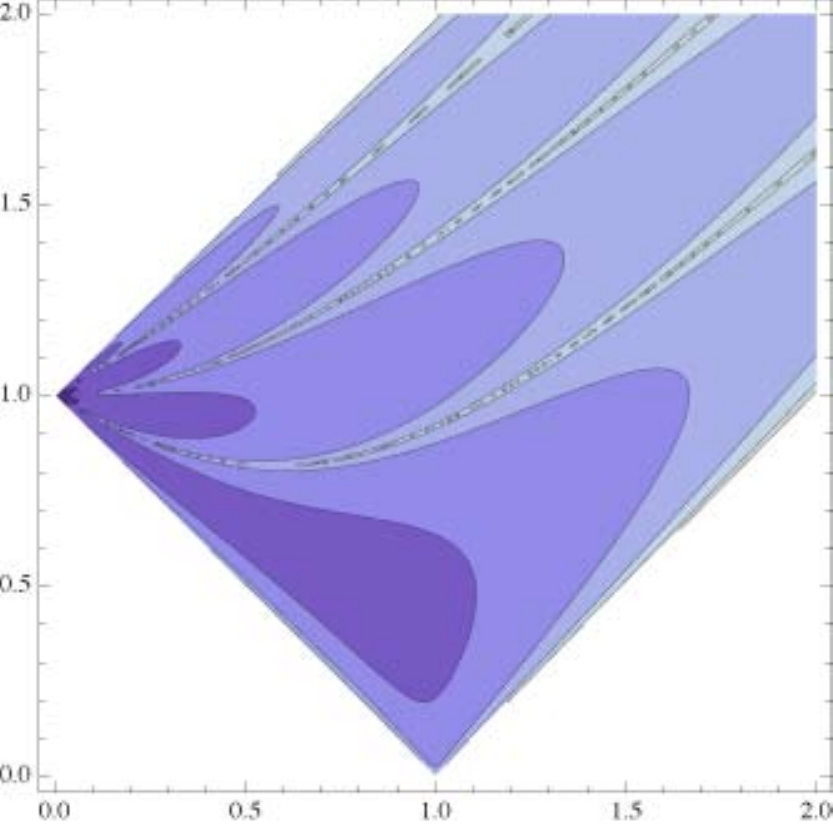}
\includegraphics[scale=0.65]{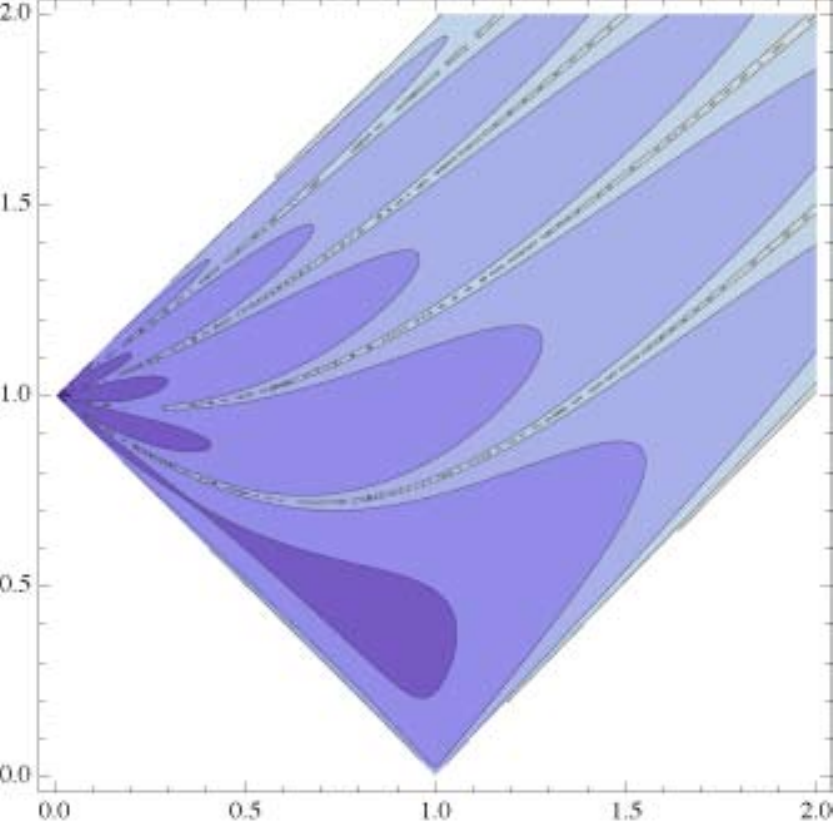}
\includegraphics[scale=0.65]{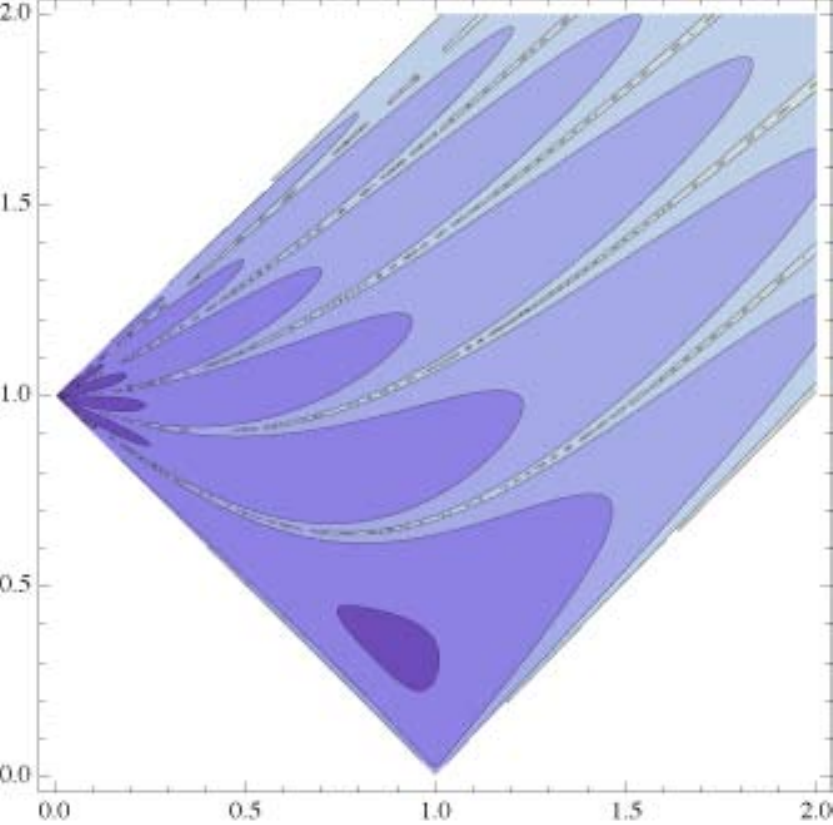}
%\par\end{centering}
\caption{ Same as above, but now we fix 
$x=1$, so the contour plots correspond to
$W^3_\ell(1,\Delta\eta;\Delta\eta')$ for the cases $\ell=$2, 3, 4 
(top panels, left to right panels), and 5, 6 and 7 (bottom panels.) 
$\Delta\eta$ corresponds to the horizontal axes, and $\Delta\eta'$
to the vertical axes.
}
\label{W3a}
\end{figure}

For two scatterings (order $\gamma^2$), the window function is
non-vanishing inside the PLC. In Figs. (\ref{W3c})-(\ref{W3a}) we show a few
examples of the window functions $W^3_\ell(x,\Delta\eta;1)$.

%%%%%%%%%%%%%%%%%%%%%%%%%%%%%%%%%%%%%%%%%%%%

\end{document}